\documentclass[aps,pre,preprint,nobibnotes,superscriptaddress,longbibliography]{revtex4-2} 
\usepackage[breaklinks,colorlinks=true,allcolors=blue]{hyperref}
\usepackage{xcolor}
\usepackage{graphicx}
\usepackage{upgreek}
\usepackage{amssymb,amsmath,mathtools}    
\usepackage{soul}
\usepackage{mathrsfs}
\usepackage[version=4]{mhchem}
\usepackage{siunitx}[=v2]
\DeclareSIUnit\Molar{M}
\begin{document}
	\title{Strong confinement of active microalgae leads to inversion of vortex flow and enhanced mixing}
	
	\author{Debasmita Mondal} 
	\affiliation{Department of Physics, Indian Institute of Science, Bangalore, Karnataka 560012, India.}
	\author{Ameya G. Prabhune}
	\affiliation{Department of Physics, Indian Institute of Science, Bangalore, Karnataka 560012, India.}
	\affiliation{Department of Physics, University of Colorado Boulder, Boulder, Colorado 80309, USA.}
	\author{Sriram Ramaswamy}	
	\affiliation{Centre for Condensed Matter Theory, Department of Physics, Indian Institute of Science, Bangalore, Karnataka 560012, India.}	
	\author{Prerna Sharma}
	\email{prerna@iisc.ac.in}
	\affiliation{Department of Physics, Indian Institute of Science, Bangalore, Karnataka 560012, India.}

\begin{abstract}
		
Microorganisms swimming through viscous fluids imprint their propulsion mechanisms in the flow fields they generate. Extreme confinement of these swimmers between rigid boundaries often arises in natural and technological contexts, yet measurements of their mechanics in this regime are absent. Here, we show that strongly confining the microalga \textit{Chlamydomonas} between two parallel plates not only inhibits its motility through contact friction with the walls but also leads, for purely mechanical reasons, to inversion of the surrounding vortex flows.
Insights from the experiment lead to a simplified theoretical description of flow fields based on a quasi-2D Brinkman approximation to the Stokes equation rather than the usual method of images. We argue that this vortex flow inversion provides the advantage of enhanced fluid mixing despite higher friction. Overall, our results offer a comprehensive framework for analyzing the collective flows of strongly confined swimmers.

\end{abstract}
	
\maketitle
	
\makeatletter
\renewcommand*{\fnum@figure}{\textbf{Figure~\thefigure}}
\makeatother
	
\clearpage \newpage

\section*{Introduction}

Fluid friction governs the functional and mechanical responses of  microorganisms which operate at low Reynolds number. They have exploited this friction and developed drag-based propulsive strategies to swim through viscous fluids \cite{Lauga2009HydroRev,Pedley1992HydroSuspensionSwimmer}. Naturally, many studies have elucidated aspects of the motility and flow fields of microswimmers in a variety of settings that mimic their natural habitats \cite{Elgeti2015Rev,Bechinger2016RevCrowdedEnvironments,SpermMicrochannel,SujitDattaNCommBacteriaPorousMedia}.
The self-propulsion of microbes in crowded and strongly confined environments is one such setting, encountered very commonly in the natural world as well as in controlled laboratory experiments. Examples include microbial biofilms, bacteria- and algae-laden porous rocks or soil \cite{QinSci2020BacterialBiofilm,AlgalBiofilmReactors,FoissnerSoilCiliates,SujitDattaNCommBacteriaPorousMedia}; parasitic infections in crowded blood streams and tissues  \cite{TrypanosomePathogenInBlood}; and biomechanics experiments using thin films and microfluidic channels \cite{DurhamSci2009ChlamyThinLayer,SpermMicrochannel,Polin2019PRLConfined,Ostapenko2018PRLConfinedChlamy,GollubPNAS2011Biomixing}. Confined microswimmers are also fundamentally interesting as active suspensions \cite{LaugaBartolo2013PRLConfined,SriramSubstratePRL} and there are efforts to mimic these by chemical and mechanical means for applications in nano- and microtechnologies \cite{Duan2015synthetic,ConfinedMicrobot2015}. 

The mechanical interaction of microswimmers with confining boundaries alters their motility and flow fields \cite{Lauga2009HydroRev,LaugaBartolo2013PRLConfined,Yeomans2016HydroMSFilm}, leading to emergent self-organization in cell-cell coordination \cite{RiedelSpermVortex2005,LibchaberPRL2DLivingActiveCrystal}, spatial distribution of cells \cite{KansoShockWavesConfinedSwim,RothschildSperm}, and ecological aspects such as energy expenditure, nutrient uptake, fluid mixing, transport and sensing \cite{LambertJFMNutrientUptakeThinFilm,Yeomans2014ConfinedStirring}. It is expected that steric interactions will dominate with increasing confinement at the swimmer-wall interface and that hydrodynamic screening by the confining wall will lead to recirculating flow patterns or vortices \cite{HowardStone2015RevBacteriaMechanics,Yeomans2016HydroMSFilm}.

Among the abundant diversity of microswimmers, the unicellular and biflagellated algae \textit{Chlamydomonas reinhardtii} (CR), with body diameter $D \approx \SI{10}{\um}$, are a versatile model system, widely used for understanding cellular processes such as carbon fixation, DNA repair and damage, phototaxis, ciliary beating \cite{ChlamyModelGrossman,Brumley2015RevAlgaEcology,Sujeet2019Phototaxis,DMSciAdvCilia2020} and physical phenomena of biological fluid dynamics \cite{Goldstein2015Rev,BrennenWinet1977CiliaRev,Rafai2010PRLEffectiveViscosity}. They are considered next-generation resources for wastewater remediation and synthesis of biofuel, biocatalysts, and pharmaceuticals \cite{AlgalBiofilmReactors,MicroalgaeFutureProspects}. Recently, extreme confinement between two hard walls has been exploited to induce stress memory in CR cells towards enhanced biomass production and cell viability \cite{Min2014CompressCRLipid,Mikulski2021CRStressMemory}. Despite the existing and emerging contexts outlined above, knowledge about how rigid walls might modify the kinetics, kinematics, fluid flow and mixing, and theoretical description of a strongly confined microalga such as CR (or any other microswimmer) is scarce. All studies prior to ours have exclusively focused on the effect of boundaries on CR dynamics in PDMS chambers or thin fluid films of height $ H \gtrsim \SI{14}{\um}$ \cite{Polin2019PRLConfined,Ostapenko2018PRLConfinedChlamy,Gollub2010PRLChlamyflowfield}, i.e., for weak confinement, $D/H < 1$.

Here, we present the first experimental measurements of the flagellar waveform, motility and flow fields of \textit{strongly confined}  CR cells placed in between two \textit{hard} glass walls $ \sim \SI{10}{\um}$ apart ($ D/H \gtrsim 1 $, denoted `H10 cells'), and infer from them the effect of confinement on kinetics, energy dissipation and fluid mixing due to the cells. We also measure the corresponding quantities for weakly confined cells placed in glass chambers of height $ H=\SI{30}{\um}$ ($ D/H \sim 0.3 $, denoted `H30 cells') for comparison. We find that the cell speed decreases significantly and the trajectory tortuosity increases with increasing confinement as we go from H30 to H10 cells.

Surprisingly, the beat-cycle averaged experimental flow field of strongly confined cells has opposite flow vorticity to that expected from the screened version of bulk flow \cite{Drescher2010ChlamyFlow,Gollub2010PRLChlamyflowfield}. This counterintuitive result comes about because the close proximity of the walls greatly suppresses the motility of the organism and, consequently, the thrust force of the flagella is balanced primarily by the non-hydrodynamic contact friction from the walls. The reason being that the flagellar thrust is largely unaffected by the walls, whereas the hydrodynamic drag on the slowly moving cell body is readily seen to be far smaller.
Understandably, theoretical predictions from the source-dipole description of strongly confined swimmers do not account for this vortex flow inversion because they include only hydrodynamic stresses \cite{LaugaBartolo2013PRLConfined,Yeomans2016HydroMSFilm}.
We complement our experimental results with a simple theoretical description of the strongly confined microswimmer flows using a quasi-2D steady Brinkman approximation to the Stokes equation \cite{brinkman1949calculation}, instead of the complicated method of recursive images using Hankel transforms \cite{LironMochon1976TwoPlates,Yeomans2016HydroMSFilm}. Solving this equation, we demonstrate that the vortex flow inversion in strong confinement is well-described as arising from a pair of like-signed force densities localized with a Gaussian spread around the approximate flagellar positions rather than the conventional three overall neutral point forces for CR \cite{Drescher2010ChlamyFlow}.
We also show that under strong confinement there is enhanced fluid transport and mixing despite higher drag due to the walls.

\clearpage

\section*{Results}

\subsection*{{\normalsize Experimental System}}

Synchronously grown wild-type CR cells (strain CC 1690) swim in a fluid medium using the characteristic breaststroke motion of two $ \sim \SI{11}{\um} $ long anterior flagella with beat frequency $ \nu_b \sim 50-60 \thinspace \si{\Hz}$. These cells are introduced into rectangular quasi-2D chambers (area, $\SI{18}{\mm} \times \SI{6}{\mm}$) made up of a glass slide and coverslip sandwich with double tape of thickness $ H = \SI{10/30}{\um}$ as spacer. Passive \SI{200}{\nm} latex microspheres are added as tracers to the cell suspension for measuring the fluid flow using particle-tracking velocimetry (PTV). We use high-speed phase-contrast imaging at $\sim  500$ frames/second and 40X magnification to capture flagellar waveform and cellular and tracer motion at a distance $ H/2 $ from the solid walls. The detailed experimental procedure is described in the Materials and Methods section.

\begin{figure}[h!]
	\centering
	\includegraphics[width=\linewidth]{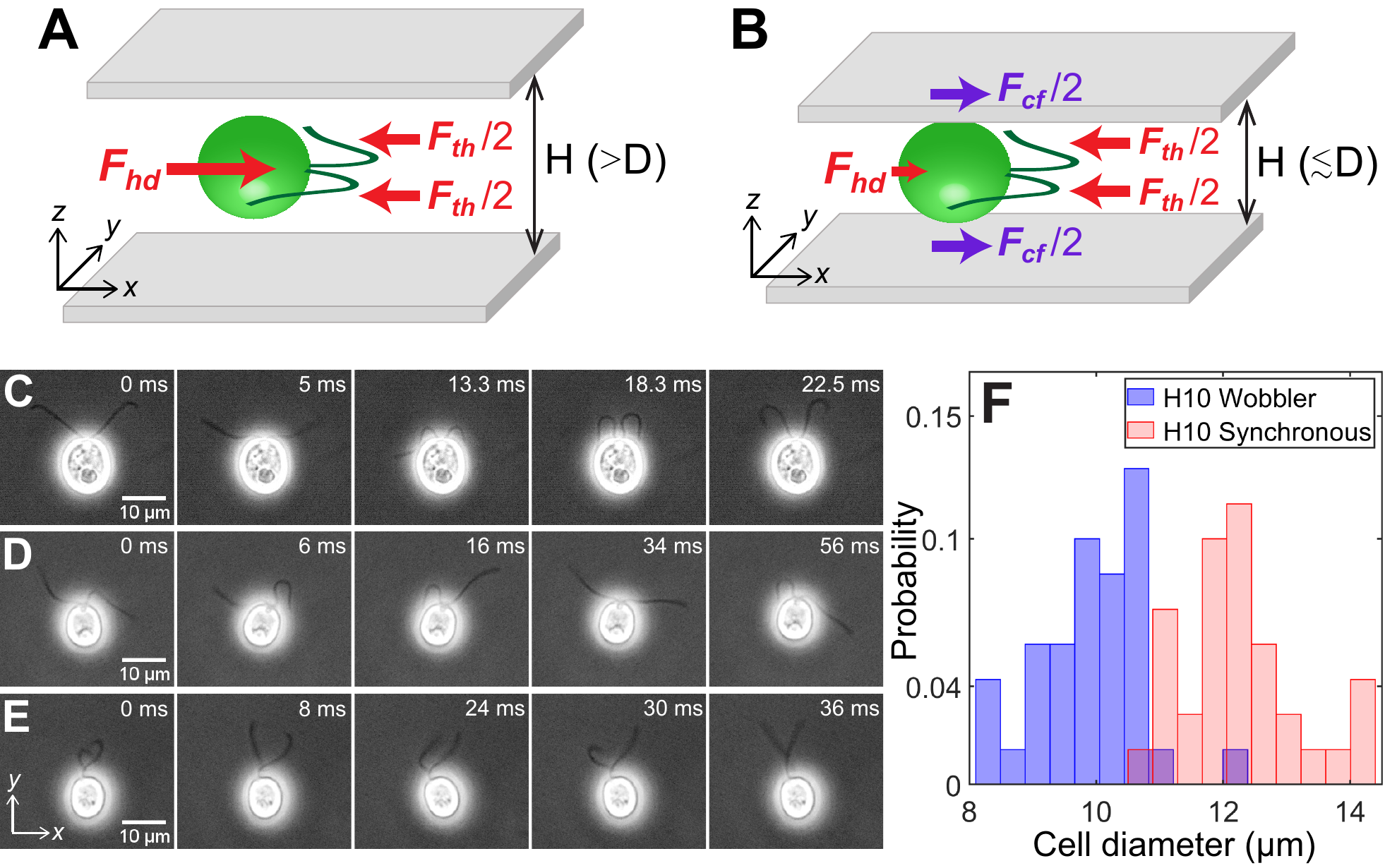}
	\caption{\textbf{Cell size affects forces acting on confined microswimmers.} Schematics of the forces exerted by a \textit{Chlamydomonas} cell (green) swimming along the $ x- $axis in between two glass plates separated by height, $ H $ under (\textbf{A}) weak confinement where the cell's body diameter, $ D<H $ and (\textbf{B}) strong confinement where $ D \gtrsim H $. Solid arrows represent local forces exerted by the cell on the surrounding medium. $ \boldsymbol{F}_{th} $ and $ \boldsymbol{F}_{hd} $ are the  propulsive thrust distributed equally between the two flagella and hydrodynamic drag due to the cell body, respectively. $\boldsymbol{F}_{\text{\textit{cf}}} $  is the contact friction with the strongly confining walls (B). Time lapse images of CR cells swimming in a quasi-2D chamber of height $ H= \SI{10}{\um} $ with (\textbf{C})~synchronously beating flagella with $ \nu_b\sim \SI{39}{\Hz} $ ($ D \sim \SI{13.2}{\um} $); (\textbf{D})~asynchronously beating flagella ($ D \sim \SI{9.9}{\um} $); and (\textbf{E})~paddler type flagellar beat ($ D \sim \SI{9.7}{\um} $). The cell bodies in (D) and (E) wobble due to their irregular flagellar beat pattern and are called `Wobblers'. (\textbf{F}) Histogram of cell body diameter in the chamber of $  H= \SI{10}{\um} $ (Number of cells, $ N=70 $). Synchronously beating cells ($ N=34 $) typically have larger diameter than Wobblers ($ N=36 $) and thus the H10 Synchronous cells with $ D/H \gtrsim 1 $ are strongly confined.}
	\label{fig:Fig1_Schematic}
\end{figure}

\subsection*{{\normalsize Mechanical equilibrium of confined cells}}

The net force and torque on microswimmers, together with the ambient medium and boundaries, can be taken to be zero as gravitational effects are negligible in the case of CR  for the range of length scales considered \cite{Drescher2010ChlamyFlow,BrennenWinet1977CiliaRev,Pedley1992HydroSuspensionSwimmer,Elgeti2015Rev,Yeomans2016HydroMSFilm}.
The two local forces exerted by any dipolar microswimmer on the surrounding fluid are flagellar propulsive thrust $\boldsymbol{F}_{th}$ and cell body drag $\boldsymbol{F}_{hd}$. They balance each other completely for any swimmer in an unbounded medium \cite{Lauga2009HydroRev,Goldstein2015Rev} and approximately in weak confinement between two hard walls (\autoref{fig:Fig1_Schematic}A). 
In these regimes, CR is the classic example of an active puller where the direction of force dipole due to thrust and drag are such that the cell draws in fluid along the propulsion axis ($ x- $axis in \autoref{fig:Fig1_Schematic}A) and ejects it in the perpendicular plane \cite{Lauga2009HydroRev}. CR is described well by three point forces or Stokeslets \cite{Drescher2010ChlamyFlow} as in \autoref{fig:Fig1_Schematic}A because the thrust is spatially extended and distributed equally between the two flagella. However, microswimmers in strong confinement between two closely spaced hard walls, $D/H \gtrsim 1$, are in a regime altogether different from bulk because the close proximity of the cells to the glass walls results in an additional drag force $\boldsymbol{F}_{\text{\textit{cf}}}$ (\autoref{fig:Fig1_Schematic}B). Therefore, the flagellar thrust is balanced by the combined drag due to the cell body and the strongly confining walls (\autoref{fig:Fig1_Schematic}B). 

\subsection*{{\normalsize Size polydispersity, confinement heterogeneity, and consequences for flagellar waveform and motility}}

We define the degree of confinement of the CR cells as the ratio $D/H$ of cell body diameter to chamber height. CR cells in chambers of height $ H=\SI{30}{\um} $ are always in weak confinement as the cell diameter varies within $ D \sim 8-14 \thinspace \si{\um} < H $. However, this dispersity in cell size becomes significant when CR cells are swimming within quasi-2D chambers of height, $ H=\SI{10}{\um} $. Here, the diameter of individual cell is crucial in determining the character -- weak or strong -- of the confinement and, as a consequence, the forces acting on the cell. Below, we illustrate how the cell size determines the type of confinement in this regime through measurements of flagellar waveform and cell motility.

CR cells confined to swim in $ H=\SI{10}{\um} $ chambers show three kinds of flagellar waveform: (a) synchronous breaststroke and planar beating of flagella interrupted by intermittent phase slips (`H10 Synchronous', \autoref{fig:Fig1_Schematic}C, Video 1); (b) asynchronous and planar flagellar beat over large time periods (\autoref{fig:Fig1_Schematic}D, Video 2); and (c) a distinctive paddling flagellar beat wherein flagella often wind around each other and paddle irregularly anterior to the cell with their beat plane oriented away from the $ x-y $ plane (\autoref{fig:Fig1_Schematic}E, Video 2). While both synchronous and asynchronous beats are typically observed for CR in bulk \cite{PolinGoldsteinScience2009} and weak confinement of \SI{30}{\um}, the paddler beat is associated with calcium-mediated mechanosensitive shock response of the flagella to the chamber walls \cite{Fujiu2011MechanoreceptionCRflagella}. The cell body wobbles for both asynchronous and paddler beat of cells (\autoref{fig:Fig1_Schematic}D \& E) and often the flagellar waveform in a single CR switches between these two kinds (Video 2). Hence, we collectively call them `H10 Wobblers' \cite{Qin2015Flagella}.

\begin{figure} [t]
	\centering
	\includegraphics[width=\linewidth]{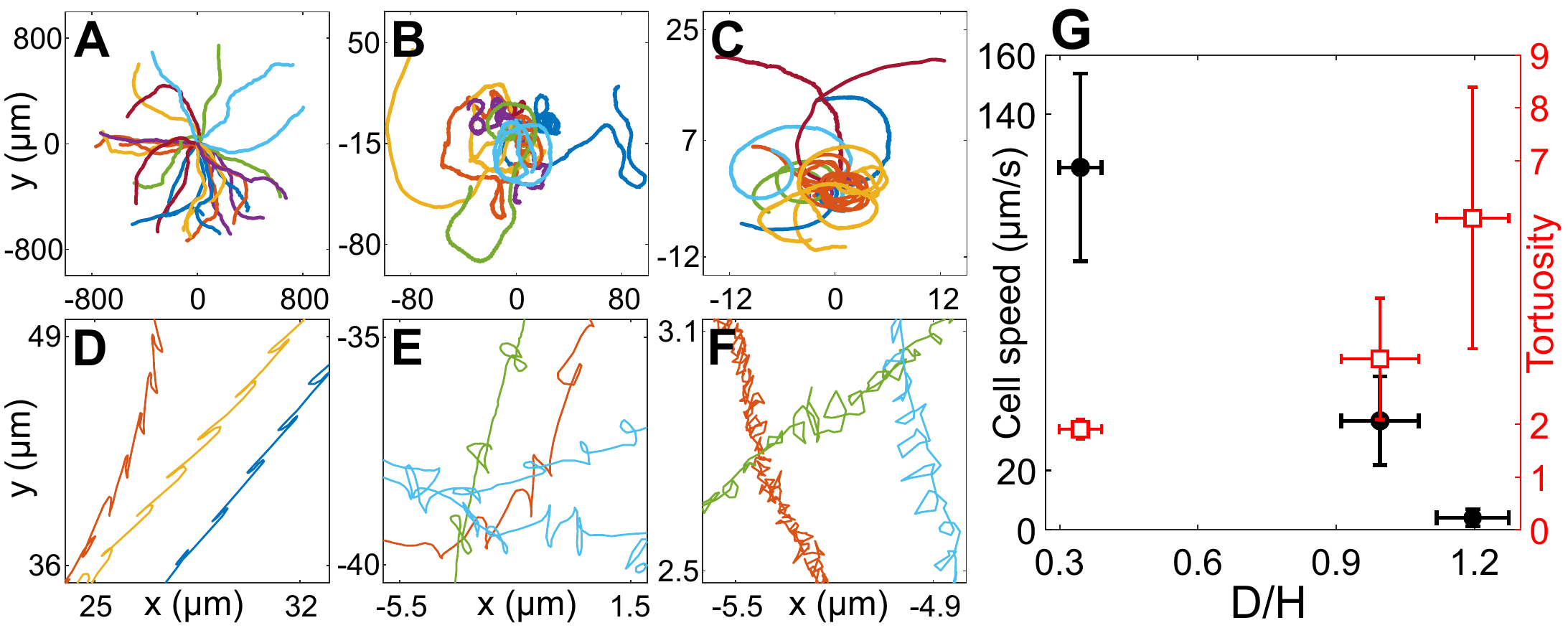} 
	\caption{\textbf{Cell motility in confinement.} Representative trajectories of CR cells in (\textbf{A}) $ H= \SI{30}{\um}$ ($ N=25 $), (\textbf{B}) $ H= \SI{10}{\um}$, Wobblers ($ N=13 $); (\textbf{C}) $ H= \SI{10}{\um}$, Synchronous cells ($ N=17 $). All of these trajectories lasted for \SI{8.2}{\second} and their initial positions are shifted to origin. (\textbf{D}), (\textbf{E}) and (\textbf{F}) are the zoomed in trajectories of (A), (B) and (C), respectively. (\textbf{G}) Cell speed (circles) and tortuosity of trajectories (squares) as a function of the degree of confinement, $ D/H $ ($ N = 52,35,23 $ for H30, H10 Wobbler and H10 Synchronous, respectively). The error bars in the plot correspond to standard deviation in diameter ($ x- $axis), cell speed and tortuosity ($ y- $axes) due to the heterogeneous population of cells.}
	\label{fig:Fig2_Motility} 
\end{figure}

We correlate the Synchronous and Wobbler nature of cells to their body diameter (\autoref{fig:Fig1_Schematic}F). The mean projected diameter in the image plane of Synchronous cells ($ D=12.28 \pm 0.94 \thinspace \si{\um} $, Number of cells, $ N = 34 $) is larger than that of Wobblers ($ D=9.92 \pm 0.85 \thinspace \si{\um}$, $ N = 36 $). Hence, the former's cell body is squished and \textit{strongly confined} in $ H=\SI{10}{\um} $ chamber in comparison with that of the latter. This leads to planar swimming of Synchronous cells, whereas Wobblers tend to spin about their body axis and trace out a near-helical trajectory which is a remnant of its behaviour in the bulk. Thus, the Wobblers likely compromise their flagellar beat into asynchrony and/or paddling over long periods, as a shock response, due to frequent mechanical interactions with the solid boundaries while rolling and yawing their cell body \cite{Fujiu2011MechanoreceptionCRflagella, Sujeet2019Phototaxis}.

The motility of CR cells in $ H=\SI{30}{\um} $ is similar to that in bulk and has the signature of back-and-forth cellular motion due to the recovery and power strokes of the flagella (\autoref{fig:Fig2_Motility}A,D). As confinement increases, the drag on the cells due to the solid walls increases and they trace out smaller distances with increasing twists and turns in the trajectory (\autoref{fig:Fig2_Motility}A-F). These phenomena can be quantitatively characterized by cell speed and trajectory tortuosity (Materials and Methods) as a function of the degree of confinement of the cells (\autoref{fig:Fig2_Motility}G). Cellular speed decreases and tortuosity of trajectories increases with increasing confinement as we go from H30 $ \to $ H10 Wobblers $ \to $ H10 Synchronous cells. Notably, the cell speed $ u $ decreases by 96\% from H30 ($ \langle u^{30} \rangle = 122.14 \pm 31.59 ~ \si{\um\per\s}$, $ N=52 $) to H10 Synchronous swimmers ($ \langle u^{10} \rangle = 4.07 \pm 2.88 ~ \si{\um\per\s} $, $ N=23 $). Henceforth, we equivalently refer to the H10 Synchronous CR as `\textit{strongly confined}' or `\textit{H10}' cells ($ D/H \gtrsim 1 $)  and the \textit{H30} cells as `\textit{weakly confined}' ($ D/H<1 $). 

We also note that the flagellar beat frequency of the strongly confined cells, $ \nu_b^{10} \approx 51.58  \pm  7.62 \thinspace \si{Hz}$ (averaged over 210 beat cycles for $ N=20 $) is  similar to that of the weakly confined ones, $ \nu_b^{30} \approx 55.27 \pm 8.22 \thinspace \si{Hz}$ (averaged over 194 beat cycles for $ N=20 $). This is because even in the $ \SI{10}{\um} $ chamber where the CR cell body is strongly confined, the flagella are beating far from the walls ($ \sim \SI{5}{\um} $) and almost unaffected by the confinement.

\subsection*{{\normalsize Experimental flow fields}}

\begin{figure}[h]
	\centering
	\includegraphics[width=0.75\linewidth]{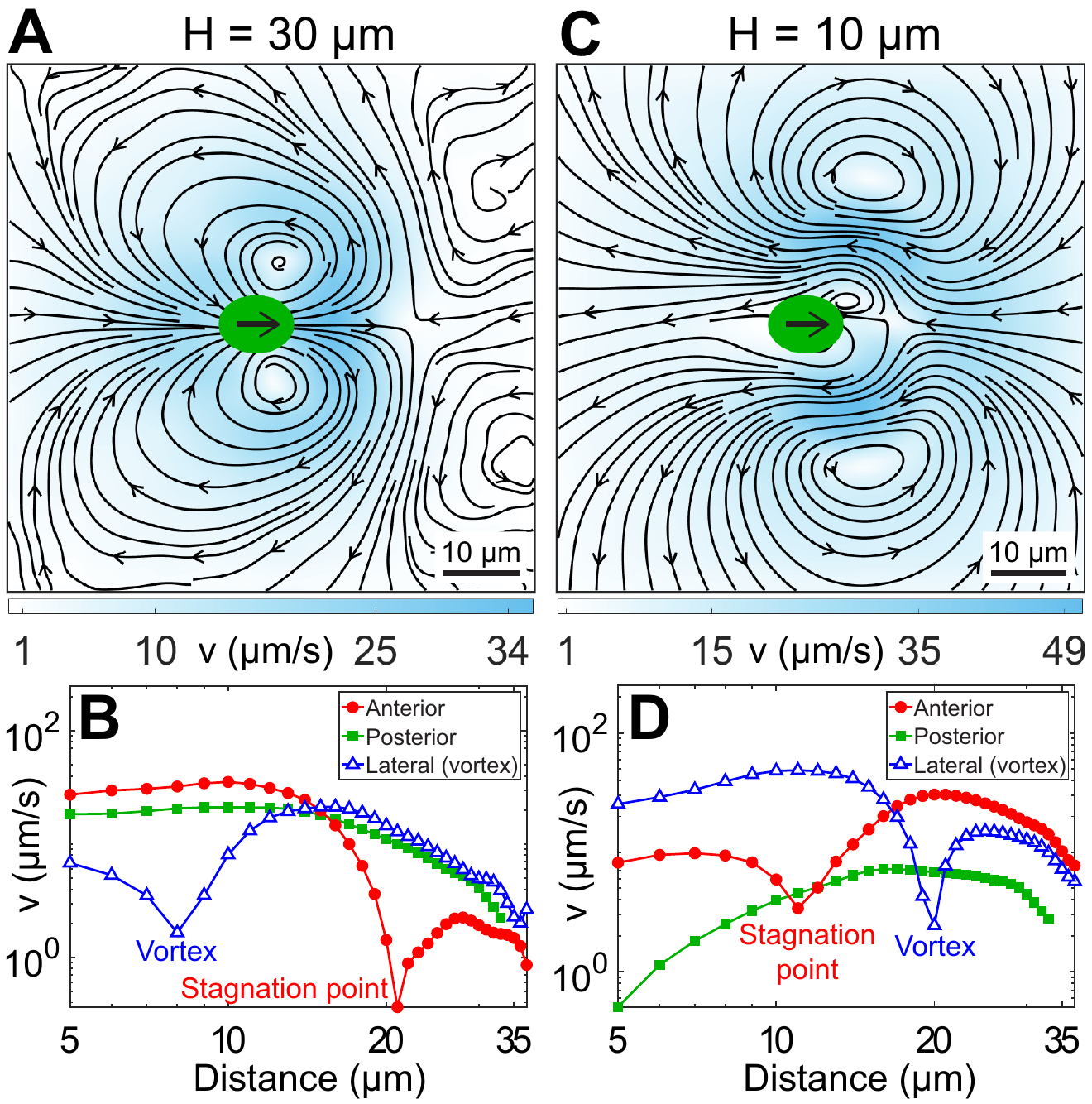} 
	\caption{\textbf{Experimental flow fields of CR cells in weak and strong confinement.} Experimentally measured, beat-averaged flow fields in the $ x-y $ plane of synchronously beating CR cells swimming in (\textbf{A}) $ H= \SI{30}{\um}$, (\textbf{C}) $ H= \SI{10}{\um}$. Black arrows on the cell body indicate that the cells are swimming to the right. Solid black lines indicate the streamlines of the flow in lab frame. The colorbars represent flow magnitude, $ v $. (\textbf{B}) and (\textbf{D}) denote the speed variation in (A) and (C), respectively, along anterior, posterior and lateral to the cell (where the vortices are present). Distances along anterior and posterior are measured along horizontal lines from the cell centre $ (0,0) $; whereas the lateral (vortex) distances are measured along the vertical line passing through $ (x,y) = (2,0) $ for (B) and $ (8,0) $ for (D), respectively.}	\label{fig:Fig3_ExptFlow} 
\end{figure}

We measure the beat-averaged flow fields of H30 and H10 CR cells to systematically understand the effect of strong confinement on the swimmer's flow field. We determine the flow field for H30 cells only when their flagellar beat is in the $ x-y $ plane (Video 3) for appropriate comparison with planar H10 swimmers. \autoref{fig:Fig3_ExptFlow}A shows the velocity field for H30 cells obtained by averaging $ \sim $ 178 beat cycles from 32 cells. It shows standard features of an unbounded CR's flow field \cite{Drescher2010ChlamyFlow,Gollub2010PRLChlamyflowfield}, namely far-field 4-lobe flow of a puller, two lateral vortices at 8-9\thinspace\si{\um} from cell's major axis and anterior flow along the swimming direction till a stagnation point, \SI{21}{\um} from the cell centre (\autoref{fig:Fig3_ExptFlow}B). These near-field flow characteristics are quite well explained theoretically by a 3-bead model \cite{Rafai2017Bead3Model,FriedrichJulicherPRL2012CR3Sphere,GolestanianPRL2013CR3SphereModel} or a 3-Stokeslet model \cite{Drescher2010ChlamyFlow}, where the thrust is distributed at approximate flagellar positions between two Stokeslets of strength ($ -1/2,-1/2 $) balanced by a  $ +1 $ Stokeslet due to viscous drag on the cell body (\autoref{fig:Fig1_Schematic}A).

The flow field of a representative H10 swimmer ($u= 5.67 \pm 1.57 ~ \si{\um\per\s}$, $ \nu_b \sim 42.67 \pm 2.24~\si{\Hz} $) is shown in \autoref{fig:Fig3_ExptFlow}C, averaged over $ \sim $ 328 beat cycles. 
Strikingly, the vortices contributing dominantly to the flow in this strongly confined geometry are opposite in sign to those in the bulk \cite{Drescher2010ChlamyFlow} or weakly confined case (H30, \autoref{fig:Fig3_ExptFlow}A).
This 2-lobed flow is distinct from expectations based on the screened version of the bulk or 3-Stokeslet flow, which is 4-lobed (\autoref{fig:Fig3_S1_ExpectedH10Flow}A). 
Importantly, the far-field flow resembles a 2D source dipole pointing opposite to the swimmer's motion, which is entirely different from that produced by the  standard source dipole theory of strongly confined swimmers (\autoref{fig:Fig3_S1_ExpectedH10Flow}B) \cite{LaugaBartolo2013PRLConfined,Yeomans2016HydroMSFilm,Polin2019PRLConfined}. This is because the source-dipole treatment does not consider the possibility  that the cells are squeezed by the walls, or in other words, it does not account for contact friction  \cite{LaugaBartolo2013PRLConfined,Yeomans2016HydroMSFilm}.
Other significant differences from the bulk flow include front-back flow asymmetry, opposite flow direction posterior to the cell, distant lateral vortices (\SI{20}{\um}) and closer stagnation point (\SI{11}{\um}) (\autoref{fig:Fig3_ExptFlow}D). All other H10 Synchronous swimmers, including the slowest ($ u \sim \SI{0.15}{\um\per\s} $) and the fastest ($ u \sim \SI{14}{\um\per\s} $) cells, show similar flow features.  Even though the flow fields of H30 and H10 cells look strikingly different, the viscous power dissipated through the flow fields is nearly the same (Appendix~\ref{App_PowerCalc}).

A close examination suggests that the vortex contents of the flow fields of \autoref{fig:Fig3_ExptFlow}A (H30) and \autoref{fig:Fig3_ExptFlow}C (H10) are mutually compatible. The large vortices flanking the rapidly moving CR in H30 are shrunken and localized close to the cell body in H10 due to the greatly reduced swimming speed. The frontal vortices generated by flagellar motion now fill most of the flow field in H10. Generated largely during the power stroke of flagella, they are opposite in sense to the vortices produced by the moving cell body.

\subsection*{{\normalsize Force balance on confined cells}}

In an unbounded fluid, the thrust $\boldsymbol{F}_{th}$ exerted by the flagellar motion of the cell balances the hydrodynamic drag $\boldsymbol{F}_{hd}$ on the moving cell body (\autoref{fig:Fig1_Schematic}A). We assume this balance holds for the case of weak confinement (H30) as well. We estimate $|\boldsymbol{F}_{hd}| =3\pi \eta D u$ as the Stokes drag on a spherical cell body of diameter $D \simeq \SI{10}{\um}$ moving at speed $u$ through a fluid of viscosity $\eta = \SI{1}{\milli\pascal\s}$ \cite{Goldstein2015Rev} which in the regime of weak confinement (H30), for a cell speed $u^{30} \approx \SI{120}{\um\per\s}$, is $\boldsymbol{F}_{hd}^{30} \approx  \SI{11.31}{\pico\newton} \thinspace \hat{\boldsymbol{x}}$, so that the corresponding thrust force $\boldsymbol{F}_{th}^{30} \approx \SI{- 11.31}{\pico\newton} \thinspace \hat{\boldsymbol{x}}$.

Given that CR operates at nearly constant thrust since $ u \propto \eta^{-1} $ \cite{Qin2015Flagella,Rafai2010PRLEffectiveViscosity} and that the flagella of the H10 cell are beating far from the walls ($ \sim \SI{5}{\um}$) with beat frequency and waveform similar to that of the H30 cell (Video 1 and Video 3), we take the flagellar thrust force in strong confinement to be $\boldsymbol{F}_{th}^{10} \approx \boldsymbol{F}_{th}^{30} \approx \SI{- 11.31}{\pico\newton} \thinspace \hat{\boldsymbol{x}}$ as in weak confinement. 
This thrust is balanced by the total drag on the cell body. The cell speed, $u^{10} \approx \SI{4}{\um\per\s}$, is down by a factor of $30$, and so is the hydrodynamic contribution to the drag if we assume the flow is the same as for the H30 geometry. Even if we take into account the tight confinement, and thus assume that the major hydrodynamic drag comes  \cite{LaugaBartolo2013PRLConfined,HowardStone2015RevBacteriaMechanics,Bhattacharya2005LubricationTwoWalls} from a lubricating film of thickness $\delta = (H-D)/2 \ll D$ between the cell and each wall, the enhancement of drag due to the fluid, logarithmic in $\delta / D$ \cite{Bhattacharya2005LubricationTwoWalls,Ganatos1980SphereBtwn2Plates}, cannot balance thrust for any plausible value of $\delta$.

The above imbalance drives the vortex flow inversion observed in \autoref{fig:Fig3_ExptFlow}C, as will be shown later theoretically, and implies that the drag is dominated by the direct frictional contact between the cell body and the strongly confining walls, which we denote by $\boldsymbol{F}_{\text{\textit{cf}}}$. Force balance on the fluid element and rigid walls enclosing the CR in strong confinement requires $\boldsymbol{F}_{th}^{10} + \boldsymbol{F}_{hd}^{10} + \boldsymbol{F}_{\text{\textit{cf}}}^{10} = 0$ (\autoref{fig:Fig1_Schematic}B). We know that the hydrodynamic drag under strong confinement is greater than \SI{0.38}{\pico\newton} (Stokes drag at $ u^{10} \approx \SI{4}{\um\per\s} $), but lack a more accurate estimate as we do not know the thickness $\delta$ of the lubricating film. We can therefore say that the contact force $\boldsymbol{F}_{\text{\textit{cf}}}^{10} \lesssim \SI{10.93}{\pico\newton} \thinspace \hat{\boldsymbol{x}}$. Thus the flagellar thrust works mainly against the non-hydrodynamic contact friction from the walls as expected due to the extremely low speed of the strongly confined swimmer.

\subsection*{{\normalsize Theoretical model of strongly confined flow}}

We begin by using the well-established far-field solution of a parallel Stokeslet between two plates by Liron \& Mochon in an attempt to explain the strongly confined CR's flow field \cite{LironMochon1976TwoPlates}. However, the theoretical flow of Liron \& Mochon decays much more rapidly than the experimental one and does not capture the vortex positions and flow variation in the experiment (Appendix~\ref{App_CompareLironMochon} and \ref{fig:FigS_A1_LironMochon}). This is because the Liron \& Mochon approximation to the confined Stokeslet flow is itself singular and also the far-field limit of the full analytical solution, so it cannot be expected to accurately explain the near-field characteristics of the experimental flow \cite{LironMochon1976TwoPlates}.

We therefore start afresh from the incompressible 3D Stokes equation, $ -\boldsymbol{\nabla} p(\boldsymbol{r}) +\eta \nabla^2 \boldsymbol{v}(\boldsymbol{r}) = 0,~ \nabla \cdot \boldsymbol{v}(\boldsymbol{r})=0$, where $ p $ and $ \boldsymbol{v} $ are the fluid pressure and velocity fields, respectively. Next, we formulate an effective 2D Stokes equation and find its point force solution. 
In a quasi-2D chamber of height $ H $, we consider an effective description of a CR swimming in the $ z=0 $ plane of the coordinate system with the first Fourier mode for the velocity profile along $z$, satisfying the no-slip boundary condition on the solid walls, $ \boldsymbol{v}(x,y,z=\pm H/2) =0 $ (\autoref{fig:Fig4_S1_FlowSchematic}). Therefore, the flow velocity varies as $  \boldsymbol{v}(x,y,z) = \boldsymbol{v}^0(x,y)\cos(\pi z/H)  $ (\autoref{fig:Fig4_S1_FlowSchematic}), where $ \boldsymbol{v}^0 = (v_x,v_y) $ is the flow profile in the swimmer's $ x-y $ plane  that is experimentally measured in \autoref{fig:Fig3_ExptFlow} \cite{FortuneGoldsteinJFM2021}. 
Substituting this form of velocity field in the Stokes equation we obtain its quasi-2D Brinkman approximation \cite{brinkman1949calculation}, which for a point force of strength $ \boldsymbol{F} $ at the $ z=0 $ plane, is
\begin{equation} \label{eq:2DStokesEqn} 
	-\boldsymbol{\nabla}_{xy} \thinspace p(\boldsymbol{r}) + \eta\bigg(\nabla_{xy}^2 -\frac{\pi^2}{H^2} \bigg) \boldsymbol{v}(\boldsymbol{r}) + \boldsymbol{F}\delta(\boldsymbol{r})= 0,\quad \nabla_{xy} \cdot \boldsymbol{v}(\boldsymbol{r})=0
\end{equation}
where $ p $ and $ \boldsymbol{v} \equiv \boldsymbol{v}_0 $ are the pressure and fluid velocity in the $ x-y $ plane and $ \nabla_{xy} =\partial_x\thinspace \boldsymbol{\widehat{x}} +\partial_y \thinspace\boldsymbol{\widehat{y}} $. We Fourier transform the above equation in 2D and invoke the orthogonal projection operator $ \boldsymbol{O}_{\boldsymbol{k}} = 1- \boldsymbol{\widehat{k}}\boldsymbol{\widehat{k}} $ to annihilate the pressure term and obtain the quasi-2D Brinkman equation in Fourier space
\begin{equation} \label{eq:2DStokesEqnFourier} 
	\boldsymbol{v}_{\boldsymbol{k}} =\frac{\boldsymbol{O}_{\boldsymbol{k}}\cdot \boldsymbol{F}}{\eta \bigg(k^2 +\dfrac{\pi^2}{H^2} \bigg)}
\end{equation}

\begin{figure}[t!]
	\centering
	\includegraphics[width=0.75\linewidth]{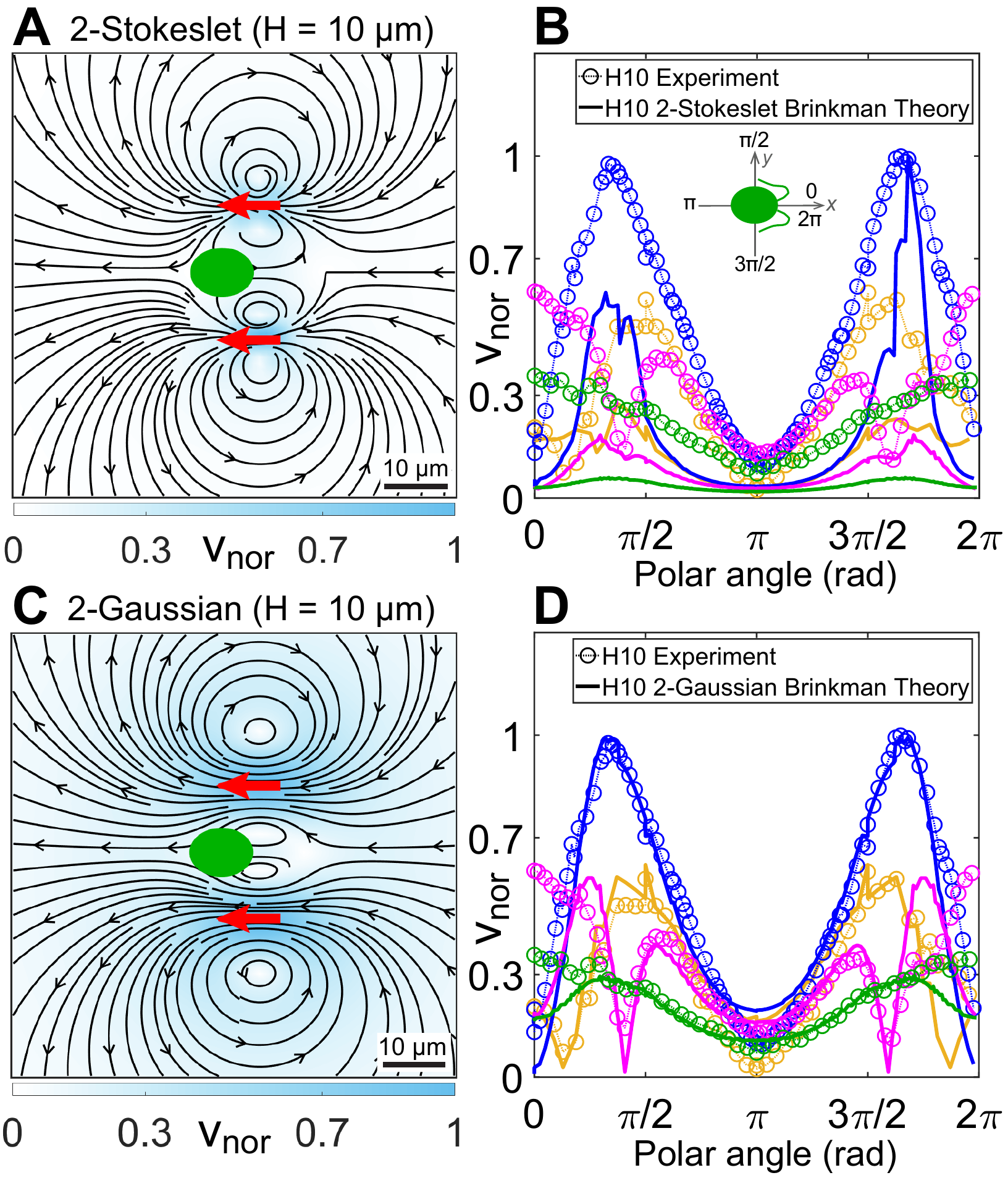} 
	\caption{\textbf{Theoretical flow fields in strong confinement.} Theoretically computed flow fields for (\textbf{A}) 2 Stokeslets and (\textbf{C}) 2-Gaussian forces, both positioned at $ (6,\pm11)\thinspace\si{\um} $ (red arrows) using the quasi-2D Brinkman equation for $ H= \SI{10}{\um}$ at the $ z=0 $ plane. The colorbars represent flow magnitudes normalised by their maximum, $v_{nor} $. (\textbf{B}) and (\textbf{D}) Comparison of normalised experimental flow of the CR in $ H= \SI{10}{\um}$  (\autoref{fig:Fig3_ExptFlow}C) with theoretical flow fields (A) and (C), respectively along representative radial distances, $ r $, from the cell centre as a function of polar angle. Inset of (B) shows the convention used for polar angle.  Plots for each $ r $ denote the flow magnitudes for those grid points which lie in the radial gap $ (r,r+1) \thinspace\si{\um}$; $ r~(\si{\um}) = 7$ (yellow), 13 (blue), 20 (magenta), 30 (green).} \label{fig:Fig4_TheoryFlow} 
\end{figure}

We perform inverse Fourier transform on \autoref{eq:2DStokesEqnFourier} in 2D for a Stokeslet oriented along the $ x $-direction, $ \boldsymbol{F} =F \thinspace \boldsymbol{\widehat{x}}$ to obtain its flow field $ \boldsymbol{v}(\boldsymbol{r}) $ at the $ z=0 $ plane (Appendix~\ref{App_IFT}). This solution is identical to the analytical closed-form expression of \cite{PushkinBees2016Brinkman}. 
We have already shown that superposing our Brinkman solution for the conventional three point forces at cell centre and flagellar positions of CR, which leads to the effective 3-Stokeslet model in 2D, is an inappropriate description of the strongly confined flow (\autoref{fig:Fig3_S1_ExpectedH10Flow}A). 
This is not surprising at this point because the force imbalance between the flagellar thrust and hydrodynamic cell drag suggests that the cell is nearly stationary compared to the
motion of its flagella.
We utilize this experimental insight by superposing only two Stokeslets of strength $  -1/2 \thinspace \boldsymbol{\widehat{x}} $ each at approximate flagellar positions $ (x_f,\pm y_f) = (6,\pm 11)\thinspace \si{\um}$ to find qualitatively similar streamlines and vortex flows (\autoref{fig:Fig4_TheoryFlow}A) as that of the experimental flow field (\autoref{fig:Fig3_ExptFlow}C). However, this theoretical `\textit{2-Stokeslet Brinkman flow}' (\autoref{fig:Fig4_TheoryFlow}A) decays faster than the experiment as shown in the quantitative comparison of these two flows in \autoref{fig:Fig4_TheoryFlow}B and \autoref{fig:Fig4_S2_H10_VxVy}, A and B. The root mean square deviation (RMSD) between these two flows in $ v_x $, $ v_y $ and $ |\boldsymbol{v}| $ are 20.3\%, 14.2\% and 22.6\%, respectively (see Materials and Methods for RMSD definition).

With the experimental streamlines and vortices well described by a 2-Stokeslet Brinkman model, we now explain the slower flow variation in experiment.  Strongly confined experimentally observed flow is mostly ascribed to the flagellar thrust, as described above. Clearly, a delta-function point force will not be adequate to describe the thrust generated by flagellar beating as they are slender rods of length $ L \sim \SI{11}{\um} $ with high aspect ratio. We, therefore, associate a 2D Gaussian source $ g(\boldsymbol{r}) = \dfrac{e^{-r^2/2\sigma^2}}{2\pi \sigma^2}  $ of standard deviation $ \sigma $, to \autoref{eq:2DStokesEqn} instead of the point-source $ \delta(\boldsymbol{r}) $, in a manner similar to the regularized Stokeslet approach \cite{Cortez2005RegularizedStokeslet3D}. Thus, the quasi-2D Brinkman equation in Fourier space (\autoref{eq:2DStokesEqnFourier}) for a Gaussian force $ \boldsymbol{F}g(\boldsymbol{r}) $ becomes,
\begin{equation} \label{eq:2DGaussEqnFourier} 
	\boldsymbol{v}_{\boldsymbol{k}} =\frac{\boldsymbol{O}_{\boldsymbol{k}}\cdot \boldsymbol{F}}{\eta \bigg(k^2 +\dfrac{\pi^2}{H^2} \bigg)} e^{-k^2\sigma^2/2} ~~.
\end{equation}
Superposing the inverse Fourier transform of the above equation for two sources of $ \boldsymbol{F} =(-1/2,-1/2) \thinspace \boldsymbol{\widehat{x}} $  at $ (x_f,\pm y_f) = (6,\pm 11) \thinspace \si{um}$ with $ \sigma \sim L/2 = \SI{5}{\um}$, we obtain the theoretical flow shown in \autoref{fig:Fig4_TheoryFlow}C. RMSD in $  v_x $, $ v_y $ and $ |\boldsymbol{v}| $ between this theoretical flow and those of the experimental one (\autoref{fig:Fig3_ExptFlow}C) are 7.8\%, 9\% and 8.3\%, respectively. Comparing these two flows along representative radial distances from the cell centre as a function of polar angle show a good agreement (\autoref{fig:Fig4_TheoryFlow}D and \autoref{fig:Fig4_S2_H10_VxVy}, C and D). Notably, \autoref{fig:Fig4_TheoryFlow}C, i.e., the `\textit{2-Gaussian Brinkman flow}', has captured the flow variation and most of the experimental flow features accurately. Specifically, these are the lateral vortices at \SI{20}{\um} and an anterior stagnation point at \SI{13}{\um}  from cell centre.
The only limitation of this theoretical model is that it cannot account for the front-back asymmetry of the strongly confined flow, as is evident from \autoref{fig:Fig4_TheoryFlow}C for the polar angles $ 0$ or $2 \pi $ and $ \pi $ which correspond to front and back of the cell, respectively. This deviation is more pronounced in the frontal region as the cell body squashed between the two solid walls mostly blocks the forward flow from reaching the cell posterior. Thus, the no-slip boundary on the cell body needs to be invoked to mimic the front-back flow asymmetry, which is a more involved analysis due to the presence of multiple boundaries and can be addressed in a follow-up study.

Now that we have explained the flow field of CR in strong confinement, we test our  quasi-2D Brinkman theory in weak confinement, $ H=\SI{30}{\um} $, where the thrust and drag forces almost balance each other. Hence, we use the conventional 3-Stokeslet model for CR, but with a Gaussian distribution for each point force. We, therefore, superpose the solution of  \autoref{eq:2DGaussEqnFourier} for 3-Gaussian forces representing the cell body and two flagella in $ H=\SI{30}{\um} $. The resulting flow field (\autoref{fig:Fig4_S3_3GaussH30}) matches qualitatively with the experimental flow field of CR in weak confinement (\autoref{fig:Fig3_ExptFlow}A). This deviation is expected in weak confinement, $ D/H \sim 0.3 $, because the quasi-2D theoretical approximation is mostly valid at $ D/H \gtrsim 1 $, even though RMSD in $ v_x $, $ v_y $ and $ |\boldsymbol{v}| $ remain in the low range at 11.4\%, 11.2\% and 13.8\%, respectively.

Together, the experimental and theoretical flow fields show that the contact friction from the walls reduces the force-dipolar swimmer in bulk or weak confinement (H30) to a force-monopole one in strong confinement (H10).

\subsection*{{\normalsize Enhancement of fluid mixing in strong confinement}}

The photosynthetic alga CR feeds on dissolved inorganic ions/molecules such as phosphate, nitrogen, ammonium, and carbon dioxide from the surrounding fluid in addition to using sunlight as the major source of energy \cite{TamHosoi2011PNASFeedingCR,Kiorboe2008PlanktonEcology}. Importantly, nitrogen and carbon are limiting macronutrients to algal growth and metabolism \cite{MicroalgaeFutureProspects,ShortGoldsteinPNASFluidTransport,Kiorboe2008PlanktonEcology}. For example, dissolved carbon dioxide in the surrounding fluid contains the carbon source essential for photosynthesis and acts as pH buffer for optimum algal growth. It is widely known that flagella-generated flow fields help in uniform distribution of these dissolved solute molecules through fluid mixing and transport which have a positive influence on the nutrient uptake of osmotrophs like CR \cite{Kiorboe2008PlanktonEcology,TamHosoi2011PNASFeedingCR,Kanso2014MixingCiliaCarpet,ShortGoldsteinPNASFluidTransport,LeptosPRLMixing3DCR,GollubPNAS2011Biomixing}. This is even more important for the strongly confined CR cells as they cannot move far enough to outrun diffusion of nutrient molecules because of slow swimming speed.

\begin{figure}[t!]
	\centering
	\includegraphics[width=\linewidth]{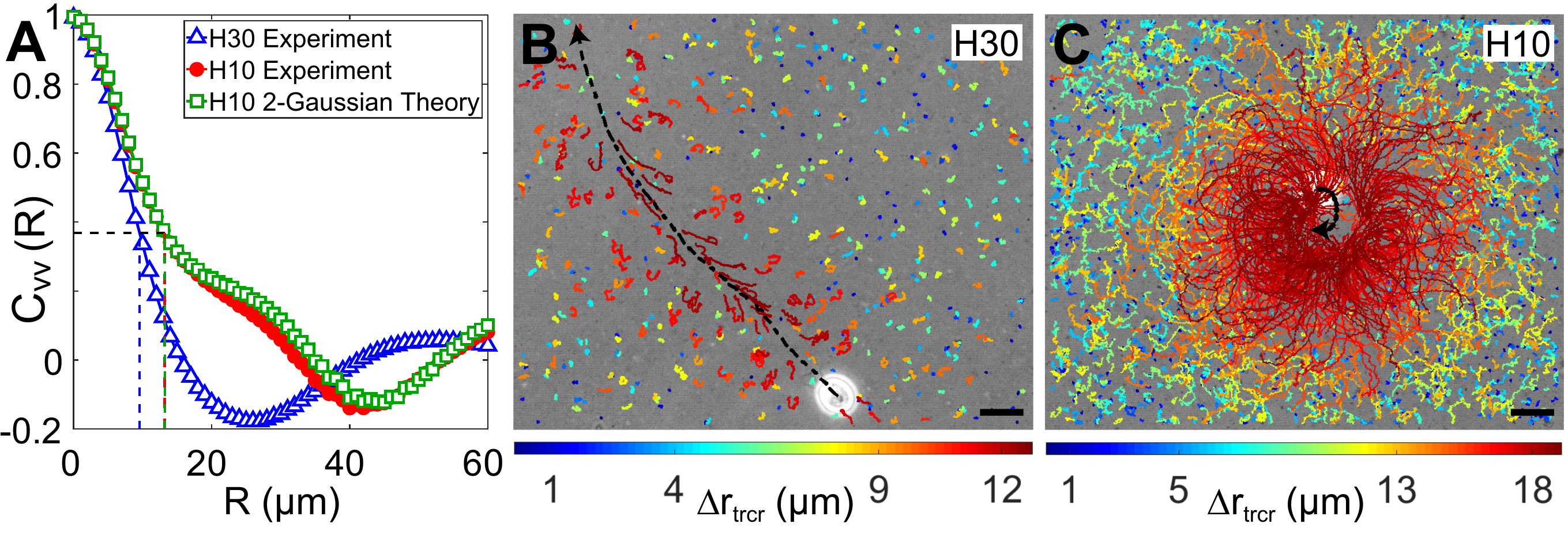} 
	\caption{\textbf{Correlation in fluid flow and tracer displacements.} (\textbf{A}) Normalised radial velocity-velocity correlation function, $ C_{vv}(R) $, of flow fields in \autoref{fig:Fig3_ExptFlow}A,C and \autoref{fig:Fig4_TheoryFlow}C. The dashed vertical lines denote the correlation length scales for the flows, $ \lambda = \SI{9.6}{\um}$~(H30) and $ \SI{13.2}{\um} $~(H10, both experiment and theory), where the correlation function decays to $ 1/e $ (horizontal dashed line). (\textbf{B}) and (\textbf{C}) Snapshots showing passive tracer trajectories (coloured) due to a CR cell (white) swimming along the black dashed arrow in $ H=\SI{30}{\um} $  and  $ H=\SI{10}{\um} $, respectively. The H30 swimmer ($ u=\SI{121}{\um\per\s} $) passes through the field of view within \SI{1.3}{\s} whereas the H10 cell ($ u=\SI{3}{\um\per\s}$) traces a semicircular trajectory staying in the field of view for the recording time of  \SI{8.2}{\s}. The tracer trajectories are colour coded, according to the colorbar below, based on their maximum displacement, $ \Delta r_{trcr} $, during a fixed lag time of $ \Delta  t = \SI{0.2}{\s} $ ($ \sim $ 10 flagellar beat cycles). Scale bars, \SI{15}{\um}.}
	\label{fig:Fig5_Mixing} 
\end{figure}

We first calculate the flow-field based P\'eclet number, $ Pe = Vl_V/D_S $ where  $ V $ and $ l_V $ are the flow-speed and diameter of the flagellar vortex, and $ D_S $ is the solute diffusivity in water, as the standard measure to characterize the relative significance of advective to diffusive transport. Using the experimentally measured flow data from \autoref{fig:Fig3_ExptFlow} and $ D_S \approx \SI{e-9}{\m^2\per\s} $  \cite{Shapiro2014PNASCoralVorticalFlow,Kiorboe2008PlanktonEcology,TamHosoi2011PNASFeedingCR}, we compute the P\'eclet numbers for the weakly and strongly confined cell to be $ Pe^{30} \approx 0.5  $ and $ Pe^{10} \approx 2 $, respectively (see \ref{tab:TableA1FlowPecletNo} and Appendix~\ref{App_Peclet}). These numbers suggest that flow-field-mediated advection does not completely dominate,  but nevertheless can play a role in nutrient uptake for small biological molecules along with diffusion-mediated transport, especially for the strongly confined cell. However, it is evident from the recorded videos of weakly and strongly confined cell suspensions that the tracers are advected more in the H10 than in the H30 chamber (Video~1 and Video~3). Hence, we attempt to quantify the observed differences in fluid mixing through correlation in flow velocity and displacement of passive tracers by the swimmers.

We calculate the normalised spatial velocity-velocity correlation function of the flow fields, $ C_{vv}(R) = \dfrac{\langle\boldsymbol{v}(r)\cdot\boldsymbol{v}(r+R)\rangle}{\langle\boldsymbol{v}(r)\cdot\boldsymbol{v}(r)\rangle} $ to estimate the enhancement of fluid mixing in strong confinement (\autoref{fig:Fig5_Mixing}A). The fluctuating flow field has a correlation length, $\lambda= \SI{13.2}{\um}$ for the strongly confined H10 flow, which is 37.5\% higher than the weakly confined flow in $ H=\SI{30}{\um} $ ($ \lambda=\SI{9.6}{\um}$), even though the cell is swimming very slowly in strong confinement.
This observation is complementary to the experiments of \cite{GollubPNAS2011Biomixing} where enhanced mixing is observed for active CR suspensions in 2D soap films compared to those in 3D unconfined fluid \cite{LeptosPRLMixing3DCR}. In their case, the reduced spatial dimension leads to long-ranged flow correlations due to the stress-free boundaries (the force-dipolar flow reduces from $ v \sim r^{-2} $ in 3D to $ v \sim r^{-1} $ in 2D).  In our case, strong confinement reduces the force-dipolar swimmer in H30 to a force-monopole one in H10 (as shown in the previous section). This leads to longer correlation length scales in the flow velocity, which implies an increased effective diffusivity (scaling, $ \sim  V_{rms}\lambda $ for a velocity field with RMS value $ V_{rms} $) of the fluid particles on time scales $ \gg \lambda/V_{rms} $, in strong confinement.

Next, we measure the displacement of the passive tracer particles when a single swimmer passes through the field of view ($\SI{179}{\um} \times \SI{143}{\um}$) in our experiments. The H30 swimmers are fast and therefore pass through this field of view in $ \sim 1-1.4 $~\si{\s} (\autoref{fig:Fig5_Mixing}B), whereas the slow-moving H10 swimmers stay in the field of view for the maximum recording time of $ \sim 8 $~\si{\s} (\autoref{fig:Fig5_Mixing}C). As the swimmer moves within the chamber, it perturbs the tracer particles. The trajectories of these tracer particles involve both Brownian components and large jumps induced by the motion and flow field of these swimmers.	We colour code the tracer trajectories based on their maximum displacement, $ \Delta r_{trcr} $, during a fixed lag time of $ \Delta t = \SI{0.2}{\s} $ ($ \sim $10 flagellar beat cycles) (\autoref{fig:Fig5_Mixing}B and C). The tracer trajectories close to the swimming path of the representative H30 swimmer (black dashed arrow) are mostly advected by the flow whereas those far away from the cell involve mostly Brownian components (\autoref{fig:Fig5_Mixing}B). However, a majority of the tracers in the full field of view are perturbed due to the H10 flow, those in the close vicinity being mostly affected (\autoref{fig:Fig5_Mixing}C). Their advective displacements are larger than that of the tracers due to H30 flow (see the colour bar below).

We define the spatial range to which a swimmer motion advects the tracers --- radius of influence, $ R_{ad} $ --- to be approximately equal to the lateral distance from the cell’s swimming path (black dashed arrow) where the tracer displacements decrease to $ \sim $ 20\% of their maxima (dark orange trajectories). The region of influence for the H30 cell is a cylinder of radius $ R_{ad} \approx \SI{15}{\um} $ with the cell’s swimming path as its axis (\autoref{fig:Fig5_Mixing}B) and that for the H10 cell is a sphere of radius $ R_{ad} \approx \SI{35}{\um} $ centred on the slow swimming cell’s trajectory (\autoref{fig:Fig5_Mixing}C). That is, the radius of influence of the H10 flow is higher than the H30 one, which corroborates the longer velocity correlation length scale in strong confinement. We also measure the mean-squared displacement (MSD) of the tracers to quantify the relative increment in the advective transport of the H10 flow with respect to the H30 one. We calculate the MSD of approximately 500 tracers in the whole field of view for each video where a single cell is passing through it and then ensemble average over 6 such videos (\autoref{fig:Fig5_S1_TracerMSD}). These plots with a scaling $ \langle \Delta r_{trcr}^2 \rangle \propto \Delta t^\alpha $ show a higher MSD exponent in H10 ($ \alpha \simeq 1.55 $) than H30 ($ \alpha \simeq 1.25 $) indicating enhanced anomalous diffusion in strong confinement. Together, \autoref{fig:Fig5_Mixing} shows that the fluid is advected more in strong confinement leading to enhanced fluid mixing and transport. In other words, the opposite vortical flows driven by flagellar beating in strong confinement help in advection-dominated dispersal of nutrients, air and \ce{CO2} in the surrounding fluid, thereby aiding the organism to avail itself of more nutrients for growth and metabolism.

\section*{Discussion}

Our results show that a prototypical puller-type of microswimmer like CR, when squeezed between two solid walls with a gap that is narrower than its size, has a remarkedly different motility and flow field from those of a bulk swimmer. In this regime of strong confinement, the cells experience a non-hydrodynamic contact friction that is large enough to decrease their swimming speed by 96\%. Consequently, their effect on the fluid is dominantly through the flagella, which pull the fluid towards the organism and therefore, the major vortices in the associated flow field have  vorticity opposite to that observed in bulk or weak confinement. This leads to an increased mixing and transport through the flow in strong confinement. These experimental results, which arise due to mechanical friction from the walls and not due to any behavioural change, establish that confinement not only alters the hydrodynamic stresses but also modifies the swimmer motility which in turn impacts the fluid flows. This coupling between confinement and motility is typically ignored in theoretical studies because the focus tends to be on the effect of confining geometry on flow-fields induced by a given set of force-generators \cite{LaugaBartolo2013PRLConfined,Yeomans2016HydroMSFilm}, which is appropriate for weak confinement, whereas strong confinement alters the complexion of forces generating the flow. Recent experimental reports have not observed the effect we discuss because they confine CR in chambers of height greater than the cell size ($ D/H \lesssim 0.7 $) \cite{Polin2019PRLConfined} where the stresses are mostly hydrodynamic  and therefore their theoretical model is force-free and different from ours (Appendix~\ref{App_CompareJeanneret}).

Our theoretical approach of using two like-signed Brinkman Stokeslets localized with a Gaussian spread on the propelling appendages can also be easily utilised to analyze flows of a dilute collection of strongly confined swimmers (Appendix~\ref{App_2Chlamy} and \ref{fig:FigS_A3_2H10Cell}).
Notably, the force-monopolar flow field of the strongly confined CR is similar to that of tethered microorganisms like \textit{Vorticella} within the slide-coverslip experimental setup \cite{Pepper2010JRSIVorticella,Malley2011PhDthesis}. Therefore, our effective 2D theoretical model involving Brinkman Stokeslet is applicable to these contexts as well. However, one needs to account for the differences in ciliary beating (two-ciliary flow for CR whereas multi-ciliated metachronal waves for \textit{Vorticella}) for a comprehensive description of the flow field closer to the organism \cite{Pepper2010JRSIVorticella,Ryu2016VorticellaRev}.

We note that even though CR is known to glide on liquid-infused solid substrates through flagella-mediated adhesive interactions \cite{ChlamyModelGrossman}, it has recently been shown that the strength of flagellar adhesion is sensitive to and switchable by ambient light \cite{Baumchen2018NatPhys}. Consequently, it is likely that CR in its natural habitat of rocks and soils would also utilise swimming in addition to gliding. Our quantitative analysis shows that despite the higher frictional drag due to the strongly confining walls, there is enhanced fluid mixing due to the H10 flow field. That is, the inverse vortical flows driven by the flagellar propulsive thrust help in advection-mediated transport of nutrients to the strongly confined microswimmer. This suggests that swimming is more efficient than gliding for CR under strong confinement (especially in low-light conditions), even though CR speeds are of the same order in both these mechanisms [$ u_{\text{glide}} \sim \SI{1}{\um\per\s} $ \cite{ChlamyModelGrossman} and $ u_{\text{swim}} \sim \SI{4}{\um\per\s} $].
We note that apart from the time-averaged flows, the oscillations produced in the flow ($ v^{osc} $) due to the periodic beating of the flagella can play a role in fluid transport and mixing for both the H30 ($ \nu_b \sim \SI{55}{Hz}$, order of magnitude estimate of $ v^{osc} \sim L \times 2\pi\nu_b \sim \SI{3450}{\um\per\s} $) and H10 ($ \nu_b \sim \SI{52}{Hz}$, $ v^{osc} \sim \SI{3270}{\um\per\s} $) cells \cite{Gollub2010PRLChlamyflowfield,KlindtFriedrich2015CRPusherPuller}.

Finally, our experimental and theoretical methodologies are completely general and can be applied to any strongly confined microswimmer, biological or synthetic from individual to collective scales. Specifically, our robust and efficient description using point or Gaussian forces in a quasi-2D Brinkman equation is simple enough to implement and analyze confined flows in a wide range of active systems.
We expect our work to inspire further studies on biomechanics and fluid mixing due to hard wall confinement of concentrated active suspensions \cite{GollubPNAS2011Biomixing,Yeomans2014ConfinedStirring,MathijssenPRLConfinedAmplifyTransport}. These effects can be exploited in realizing autonomous motion through microchannel for  biomedical applications and in microfluidic devices
for efficient control, navigation and trapping of microbes and synthetic swimmers \cite{ParkDrugDelivery,KarimiTrapMicrofluidics,ConfinedMicrobot2015}. 

\vskip 20mm

\section*{Materials and Methods}

\subsection*{Surface modification of microspheres and glass surfaces} 

CR cells are synchronously grown in 12:12 hour light:dark cycle in Tris-Acetate-Phosphate (TAP+P) medium. This culture medium contains divalent ions such as \ce{Ca^2+}, \ce{Mg^2+}, \ce{SO4^2-} which decrease the screening length of the \SI{200}{\nm} negatively charged microspheres, thereby promoting inter-particle aggregation and sticking to glass surfaces and CR's flagella. Therefore, the sulfate latex microspheres (S37491, Thermo Scientific) are sterically stabilised by grafting long polymer chains of polyethylene glycol (mPEG-SVA-20k, NANOCS, USA) with the help of a positively charged poly-L-lysine backbone (P7890, 15-30kDa, Sigma)  \cite{DMSciAdvCilia2020}. In addition, the coverslip and slide surfaces are also cleaned and coated with polyacrylamide brush to prevent non-specific adhesion of microspheres and flagella to the glass surfaces, prior to sample injection \cite{DMSciAdvCilia2020}.

\subsection*{Sample imaging} 

Cell suspension is collected in the logarithmic growth phase within the first 2-3 hours of light cycle and re-suspended in fresh TAP+P medium. After 30 minutes of equilibration, the cells are injected into the sample chamber. The sample chamber containing cells and tracers is mounted on an inverted microscope (Olympus IX83/IX73) and placed under red light illumination ($ > $ 610 nm) to prevent adhesion of flagella \cite{Baumchen2018NatPhys} and phototactic response of CR \cite{Sineshchekov2002PNAS}. We let the system acclimatize in this condition for 40 minutes before recording any data. All flow field data, flagellar waveform and cellular trajectory (except for \autoref{fig:Fig2_Motility}A) are captured using a 40X phase objective (Olympus, 0.65 NA, Plan N, Ph2) coupled to a high speed CMOS camera (Phantom Miro C110, Vision Research, pixel size = $ \SI{5.6}{\um}$) at 500 frames/s. As CR cells move faster in $ H=\SI{30}{\um} $ chamber, a  8.2 second long trajectory cannot be captured at that magnification. So we used a 10X objective in bright field (Olympus, 0.25 NA, PlanC N) connected to a high speed camera of higher pixel length (pco.1200hs, pixel size = $ \SI{12}{\um} $) at 100 frames/s to capture 8.2 s long trajectories of H30 cells (\autoref{fig:Fig2_Motility}A).

Our observations are consistent across CR cultures grown on different days and cultures inoculated from different colonies of CR agar plates. We have prepared at least 15-18 samples of dilute CR suspensions from 8 different days/batches of cultures, each for chambers of height 10 and 30 \si{\um}. Our imaging parameters remain same for all observations. We also use the same code, which is verified from standard particle tracking videos, for tracking all the cells. We modify the cell tracking code to track the tracer motion for calculating the flow-field data.

\subsection*{Height measurement of sample chamber} 
We use commercially available double tapes of thickness 10 and 30 \si{\um} (Nitto Denko Corporation) as spacer between the glass slide and coverslip. To measure the actual separation between these two surfaces, we stick 200 nm microspheres to a small strip ($\SI{18}{\mm} \times \SI{6}{\mm}$) on both the glass surfaces by heating a dilute solution of microspheres. Next, we inject immersion oil inside the sample chamber to prevent geometric distortion due to refractive index mismatch between objective immersion medium and sample. The chamber height is then measured by focusing the stuck microspheres on both surfaces through a 60X oil-immersion phase objective (Olympus, 1.25 NA). We find the measured chamber height  for the 10 \si{\um} spacer to be $ 10.88\pm 0.68 \thinspace \si{\um}$ and for the 30  \si{\um} spacer to be $ 30.32\pm 0.87\thinspace \si{\um}$, from 8 different samples in each case.

\subsection*{Particle Tracking Velocimetry (PTV)} 
The edge of a CR cell body appears as a dark line (\autoref{fig:Fig1_Schematic}C to E) in phase contrast microscopy and is detected using ridge detection in ImageJ \cite{RidgeDetectPlugin}. An ellipse is fitted to the pixelated CR's edge and the major axis vertex in between the two flagella is identified through custom-written MATLAB codes (refer to source code file). The cell body is masked and the tracers' displacement in between two frames (time gap, 2 ms) are calculated in the lab frame using standard MATLAB tracking routines \cite{Matlabtrack}. The velocity vectors obtained from multiple beat cycles are translated and rotated to  a common coordinate system where the cell's major axis vertex is pointing to the right (\autoref{fig:Fig3_ExptFlow}A,C). Outliers with velocity magnitude more than six standard deviations from the mean are deleted.  The resulting velocity vectors from all beat cycles (including those from different cells in $ H=\SI{30}{\um}$) are then placed on a mesh grid of size $\SI{2.24}{\um} \times \SI{2.24}{\um}$ and the mean at each grid point is computed. The gridded velocity vectors are then smoothened using a $ 5 \times 5 $ averaging filter. Furthermore, for comparison with theoretical flow, the $ x $ and $ y $ components of the velocity vectors are interpolated on a grid size of $ 1 \times 1 \thinspace \si{\um^2}$. Streamlines are plotted using the `\textit{streamslice}' function in MATLAB.

\subsection*{Trajectory tortuosity} 
Tortuosity characterizes the number of twists or loops in a cell's trajectory. It is given by the ratio of arclength to end-to-end distance between two points in a trajectory. We divide each trajectory into segments of arc-length $ \approx \SI{20}{\um}$. We calculate the tortuosity for individual segments and find their mean for each trajectory.
We consider the trajectories of all cells whose mean speed $> \SI{1}{\um\per\s}$ and are imaged at 500 frames/s through 40X objective for consistency. There were 52 H30 cells, 35 H10 Wobblers and 23 H10 Synchronous cells which satisfied these conditions and the data from these cells constitute \autoref{fig:Fig2_Motility}G. 

\subsection*{Root Mean Square Deviation (RMSD)} 

The match between experimental and theoretical flow fields is quantified by the root-mean-square deviation (RMSD) of their velocities in the normalised scale ($ v/v_{\text{max}} $). $RMSD = \sqrt{\sum_{j=1}^{NG} (v^{\text{expt}}_j - v^{\text{th}}_j)^2/NG}$, where $v^{\text{expt}}_j$ and $v^{\text{th}}_j$ are the experimental and theoretical values of the velocity fields at the $j$-th grid point, respectively, and $NG$ is the total number of grid points. We calculate RMSD in the $ x $ and $ y $ components of the flow velocity i.e., in $ v_x $ and $ v_y $, respectively, for  a comparison of the vector nature of the flow fields. This is because  the signed magnitudes of $ v_x $ and $ v_y $ determine the vector direction of the flow. We also calculate RMSD in the flow speed ($ |\boldsymbol{v}|=[v_x^2+v_y^2]^{1/2} $) to compare their scalar magnitudes.


\vskip 10mm

\textbf{Acknowledgements}: We acknowledge Aparna Baskaran, Ramin Golestanian, Ayantika Khanra, Swapnil J. Kole, Malay Pal, Balachandra Suri and Ronojoy Adhikari for useful discussions. 

\textbf{Funding}: This work is supported by the DBT/Wellcome Trust India Alliance Fellowship [grant number IA/I/16/1/502356] awarded to P.S. S.R. acknowledges support from a J C Bose Fellowship of the SERB (India) and from the Tata Education and Development Trust.

\textbf{Author Contributions}: D.M. and P.S. conceived and designed the experiments. S.R. proposed the theoretical model. D.M. performed experiments, data analysis and theoretical computations. A.G.P. carried out theoretical calculations at the initial stages of this work. D.M., P.S. and S.R. interpreted the experiments and wrote the manuscript.

\textbf{Competing financial interests}: The authors declare no competing financial interests.

\vskip 10mm


\bibliography{RefChlamy2d}

\vskip 10mm


\clearpage \newpage
\section*{Figure supplements}

\renewcommand{\thefigure}{3---figure supplement~1}
\begin{figure}[h]
	\centering 
	\includegraphics[width=0.78\linewidth]{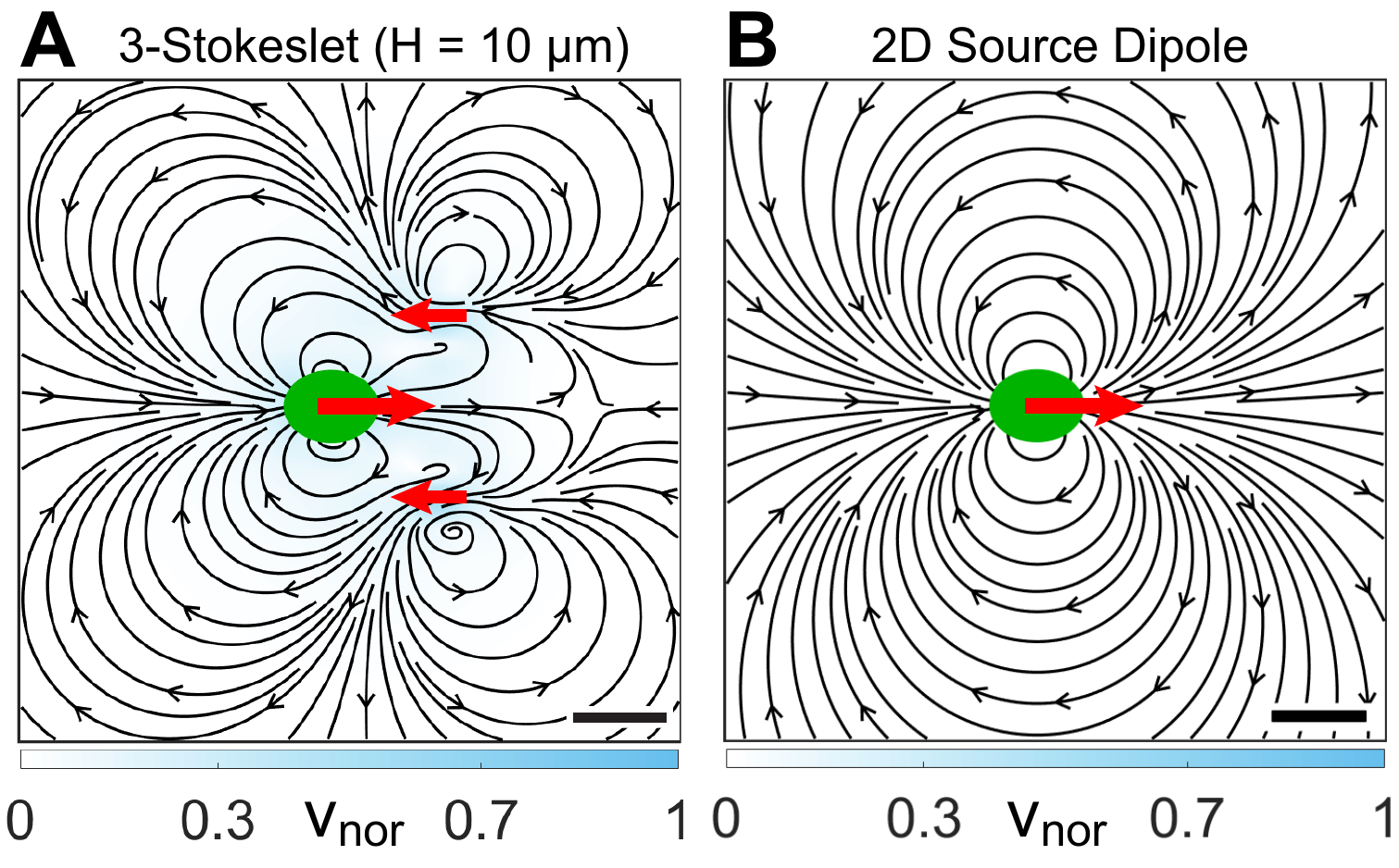}
	\caption{ \textbf{Expected flow fields of a strongly confined CR using conventional theoretical approaches.}  (\textbf{A}) Theoretically computed \textit{near-field flow} characteristics as expected from the screened version of the bulk flow field i.e., from the 3-Stokeslet model in $ H= \SI{10}{\um}$. The 3 Stokeslets denoted by red arrows represent the cell drag of strength $ +1 $ at (0,0) and flagellar thrust of strength $ -1/2 $ each at (12,$ \pm $10)\thinspace\si{\um}. This flow field is calculated using the quasi-2D Brinkman equation, which is introduced later in this article.  (\textbf{B}) Theoretically predicted \textit{far-field flow} of a microswimmer in confinement (but under the influence of hydrodynamic stresses only) which is that of a 2D source dipole oriented along the propulsion direction (denoted by red arrow) \cite{Yeomans2016HydroMSFilm,LaugaBartolo2013PRLConfined}. The colorbars represent flow magnitudes normalised by their maximum, $v_{nor} $. Scale bars, \SI{10}{\um}.}
	\label{fig:Fig3_S1_ExpectedH10Flow}
\end{figure}

\renewcommand{\thefigure}{4---figure supplement~1}
\begin{figure}[h]
	\centering
	\includegraphics[width=0.43\linewidth]{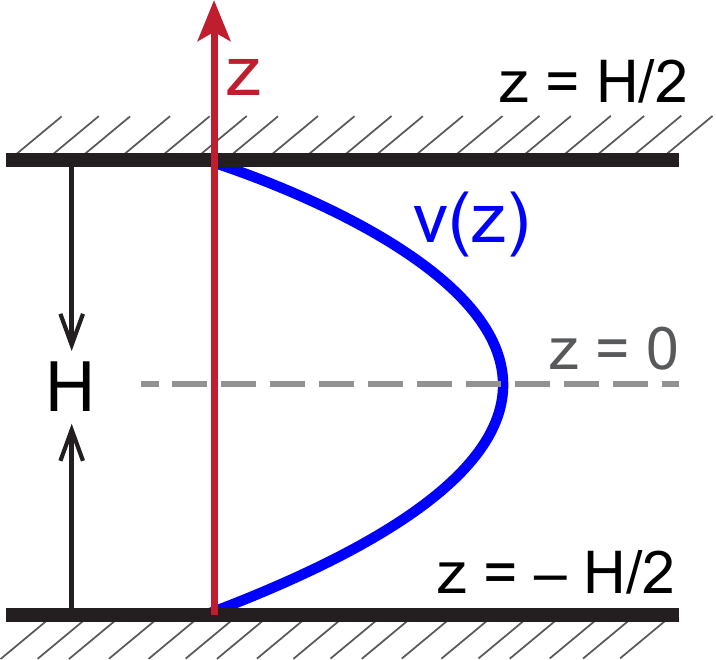}
	\caption{\textbf{Schematic of velocity profile along the confining direction.} Schematic of flow profile along $ z $-direction, $ v(z) \sim \cos(\pi z/H) $, in a chamber of height $ H $ bounded by two solid walls.}\label{fig:Fig4_S1_FlowSchematic}
\end{figure}

\renewcommand{\thefigure}{4---figure supplement~2}
\begin{figure}[h]
	\centering
	\includegraphics[width=0.82\linewidth]{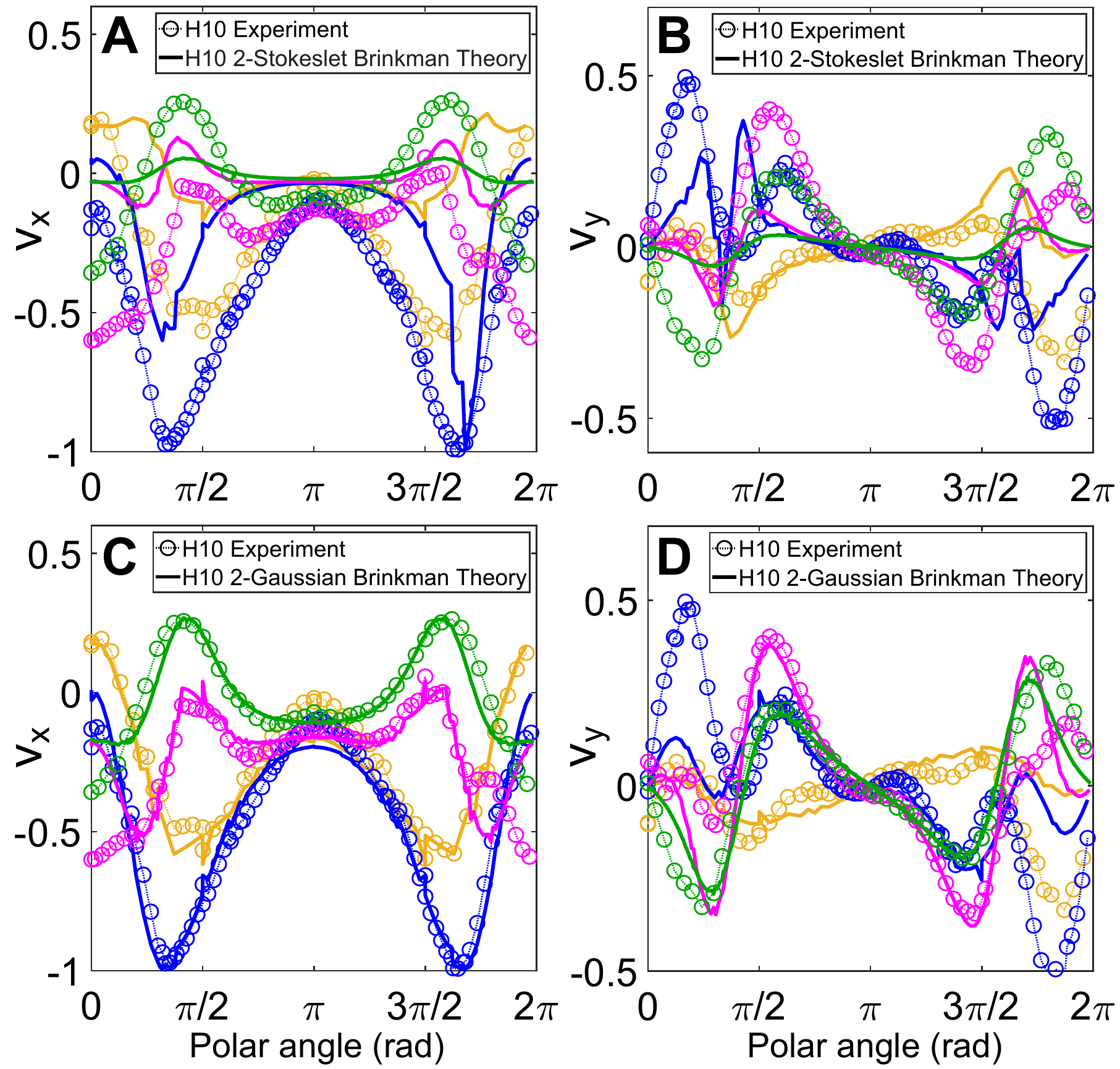}
	\caption{\textbf{Comparison in the direction of flow fields between experiment and theory.} Comparison of (\textbf{A}, \textbf{C}) $ v_x $  and (\textbf{B}, \textbf{D}) $ v_y $  between normalised experimental flow of the CR in $ H= \SI{10}{\um}$  (\autoref{fig:Fig3_ExptFlow}C) with theoretical flow fields (\autoref{fig:Fig4_TheoryFlow}A and \autoref{fig:Fig4_TheoryFlow}C), respectively, along representative radial distances, $ r $, from the cell centre as a function of polar angle. Plots for each $ r $ denote the flow components for those grid points which lie in the radial gap $ (r,r+1) \thinspace\si{\um}$; $ r~(\si{\um}) = 7$ (yellow), 13 (blue), 20 (magenta), 30 (green).}\label{fig:Fig4_S2_H10_VxVy}
\end{figure}

\renewcommand{\thefigure}{4---figure supplement~3}
\begin{figure}[h]
	\centering
	\includegraphics[width=0.81\linewidth]{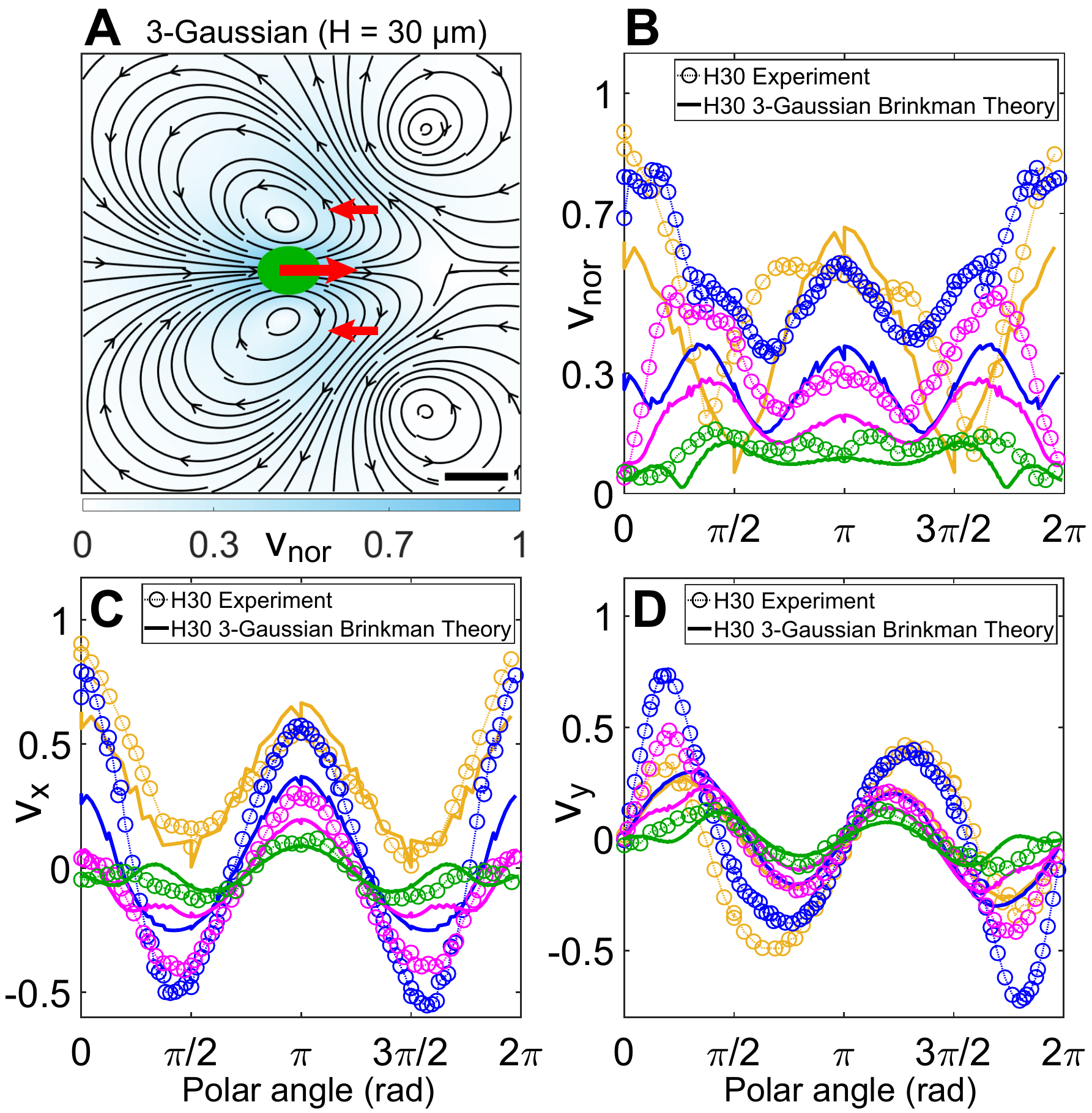}
	\caption{\textbf{Theoretical flow field in weak confinement.} (\textbf{A}) Theoretically computed flow field from 3-Gaussian forces (cell drag of strength $ +1 $ at (0,0), flagellar thrust of strength $ -1/2 $ each at (12,$ \pm $10)\thinspace\si{\um}; all denoted by red arrows) using the quasi-2D Brinkman equation for $ H= \SI{30}{\um}$ at the $ z=0 $ plane. The Gaussian standard deviation, $ \sigma $ for cell and flagellum are 3 and 5\thinspace\si{\um}, respectively. The colorbar represents flow magnitude normalised by its maximum, $v_{nor} $. Scale bar, \SI{10}{\um}.  Comparison of (\textbf{B}) $ |\boldsymbol{v}| $, (\textbf{C}) $ v_x $  and (\textbf{D}) $ v_y $ between normalised experimental (\autoref{fig:Fig3_ExptFlow}A) and theoretical flow field (A) of a cell swimming in $ H= \SI{30}{\um}$ along representative radial distances, $ r $, from the cell centre as a function of polar angle.  The convention used for polar angle is same as in \autoref{fig:Fig4_TheoryFlow}B/D. Plots for each $ r $ denote the flow magnitudes for those grid points which lie in the radial gap $ (r,r+1)\thinspace \si{\um}$; $ r~(\si{\um}) = 7$ (yellow), 13 (blue), 20 (magenta), 30 (green).}\label{fig:Fig4_S3_3GaussH30}
\end{figure}

\renewcommand{\thefigure}{5---figure supplement~1}
\begin{figure}[h]
	\centering
	\includegraphics[width=0.88\linewidth]{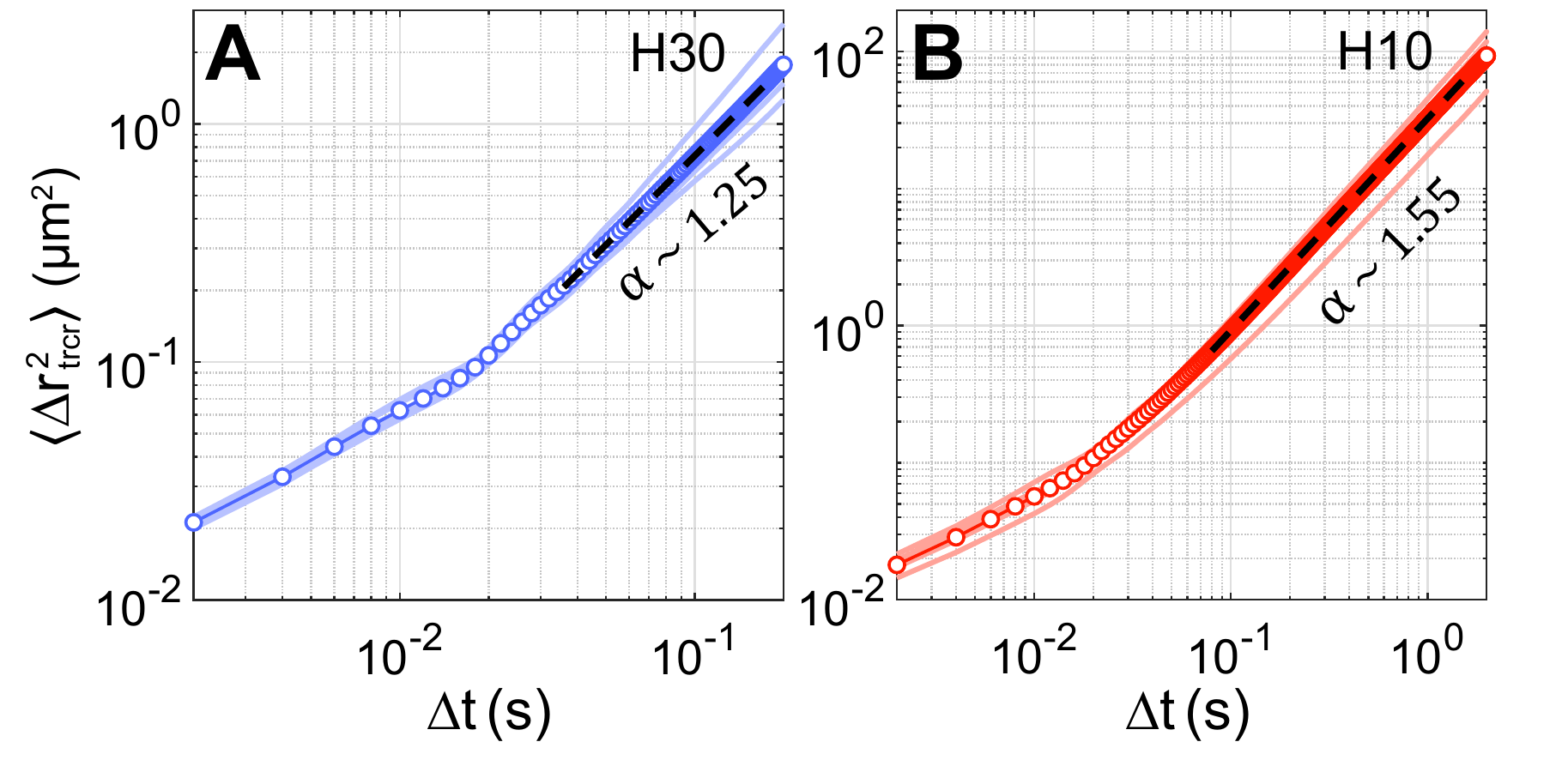}
	\caption{\textbf{Mean squared displacement (MSD) of tracers.} MSD of tracers, $ \langle \Delta r_{trcr}^2 \rangle$, due to a representative CR cell swimming through the field of view in (\textbf{A}) $ H = \SI{30}{\um} $ and (\textbf{B}) $ H= \SI{10}{\um} $. Semi-transparent lines represent the MSD of $ \sim $ 500 tracers for each video where a single swimmer is passing through the field of view and the solid line with symbols denotes the average tracer MSD from 6 such videos. Black dashed lines indicate linear fit to the log-log data of average MSD vs lag-time ($ \Delta t $) where $ \alpha $ denotes the MSD exponent, $ \langle \Delta r_{trcr}^2 \rangle \propto \Delta t^\alpha $.} \label{fig:Fig5_S1_TracerMSD}
\end{figure}


\clearpage \newpage
\section*{Video Captions}

\textbf{VIDEO 1: Video of a strongly confined Chlamydomonas cell swimming with synchronous beat in presence of tracers.} High-speed video microscopy of a strongly confined swimmer (synchronously beating \textit{Chlamydomonas} cell in $ H=\SI{10}{\um}$ chamber) in presence of tracer particles at 500 frames/s. This phase-contrast video clearly shows the synchronous breaststroke and planar beating of flagella with intermittent phase slips. This is the representative cell whose flow field is shown in \autoref{fig:Fig3_ExptFlow}C. The direction of vortex flow is evident from the tracers' motion. 

\vspace{0.4cm}

\textbf{VIDEO 2: Video of wobbling Chlamydomonas cells with asynchronous or paddling flagellar beat.} Flagellar waveform of \textit{Chlamydomonas} cells in $ H=\SI{10}{\um}$ chamber with wobbling cell body i.e., H10 Wobblers. The video is divided into 3 parts. The first part shows the asynchronous and planar flagellar beat of a cell which leads to a wobbling motion of the cell body. The second part shows the distinctive paddling flagellar beat of a cell, anterior to the cell body. Here, the flagellar beat plane is perpendicular to the imaging $ x-y $ plane and one of the flagella is mostly out of focus. In both these cases, the cell bodies wobble due to their irregular flagellar beat pattern. The third part shows a representative H10 Wobbler which switches from paddling beat to an asynchronous one. 

\vspace{0.4cm}

\textbf{VIDEO 3: Video of a weakly confined Chlamydomonas cell swimming in presence of tracers.} High-speed video microscopy of a weakly confined \textit{Chlamydomonas} cell swimming in $ H=\SI{30}{\um} $ chamber in presence of tracer particles at 500 frames/s. This video shows the natural motility of cells in bulk where they spin about their body axis. The video starts with the cell and its flagella beating in the image plane. At $ \sim  90 - 180 \thinspace\si{\ms}$, the flagellar beat of the cell is out of the image plane, when the cell body is rotating about its axis. The flow field is calculated only when the flagellar beat of the H30 cell is in the image plane, i.e. for $ 0 - 90 \thinspace\si{\ms} $ and $ 180 - 252 \thinspace\si{\ms} $ for this particular video.


\clearpage \newpage

\section*{Appendix 1}

\makeatletter
\renewcommand*{\fnum@figure}{\textbf{\thefigure}}
\renewcommand*{\fnum@table}{\textbf{\thetable}}
\makeatother

\newcommand{\beginappendix}{
	\setcounter{equation}{0}
	\renewcommand{\theequation}{A\arabic{equation}}
	\renewcommand{\thetable}{Appendix 1---table \arabic{table}}
}

\beginappendix

\renewcommand\thesubsection{1.\arabic{subsection}}

\subsection{{\normalsize Power dissipated through the flow fields}} \label{App_PowerCalc}

In low-Reynolds-number flows, the power $P$ generated by a microswimmer is dissipated through the induced flow fields as $ P = 2\eta \int_V (\boldsymbol{\Gamma}:\boldsymbol{\Gamma})\thinspace dV $ \cite{Gollub2010PRLChlamyflowfield}. Here, $ \eta $ is the fluid viscosity, $ \boldsymbol{\Gamma} = \frac{1}{2} [\boldsymbol{\nabla v}+(\boldsymbol{\nabla v})^T] $ is the fluid strain rate due to gradients in the flow velocity $ \boldsymbol{v} $, and the integral is over the quasi-2D chamber of height $ H $. Roughly, for flows in bulk or in 2D fluid films, the velocity gradient along the chamber height is negligible and only the $2 \times 2$ part of $\boldsymbol{\Gamma}$ corresponding to directions in the plane perpendicular to the confinement direction has non-negligible components \cite{Gollub2010PRLChlamyflowfield}. This is not true in our case because the rigid boundaries act as momentum sinks, imposing a significant gradient in the fluid flow along the confinement direction $ z $. Since the flow velocity varies as $  \boldsymbol{v}(x,y,z) = \boldsymbol{v}^0(x,y)\cos(\pi z/H)  $ (refer to \autoref{fig:Fig4_S1_FlowSchematic} and associated main text), the norm-squared strain rate tensor for hard-wall confined flows is given by $\boldsymbol{\Gamma}:\boldsymbol{\Gamma}= (\boldsymbol{\Gamma}:\boldsymbol{\Gamma})^{\text{bulk}} + \dfrac{(\pi v^0)^2}{2H^2} \sin^2 \bigg(\dfrac{\pi z}{H} \bigg)$ where $ (\boldsymbol{\Gamma}:\boldsymbol{\Gamma})^{\text{bulk}} =(\partial_xv_x)^2 + \frac{1}{2}(\partial_yv_x+\partial_xv_y)^2 + (\partial_yv_y)^2 $ and $ \boldsymbol{v}^0 = (v_x,v_y) $ is the flow profile in the swimmer's $ x-y $ plane  that is experimentally measured in \autoref{fig:Fig3_ExptFlow}. We calculate the viscous power dissipation from the beat-averaged flow fields of CR to be $ P^{30} = \SI{0.78}{fW} $ in weak confinement and $P^{10}=\SI{1.05}{fW}$ in strong confinement. These values are of the same order for both types of confinement and also to that measured for CR in  thin fluid films ($ \text{P}_\text{mean flow} $ in Fig. 4a of \cite{Gollub2010PRLChlamyflowfield}).

\subsection{{\normalsize Comparison of our experimental flow data in strong confinement with Liron \& Mochon's theoretical solution}} \label{App_CompareLironMochon}

The far-field solution of Liron \& Mochon for a parallel Stokeslet, $ \boldsymbol{F} $ located midway between two no-slip plates is given by $ v_i^{LM}(r) = Q^{SD} \bigg(-\frac{\delta_{ij}}{r^2} + \frac{2r_ir_j}{r^4}\bigg)F_j $, which is equivalent to that of a 2D source dipole of strength $ Q^{SD} = \frac{3H}{8\pi\eta} \frac{z}{H} \big(1-\frac{z}{H} \big)$ \cite{LironMochon1976TwoPlates}. 

\renewcommand{\thefigure}{Appendix 1---figure~1}
\begin{figure}[h]
	\centering
	\includegraphics[width=\linewidth]{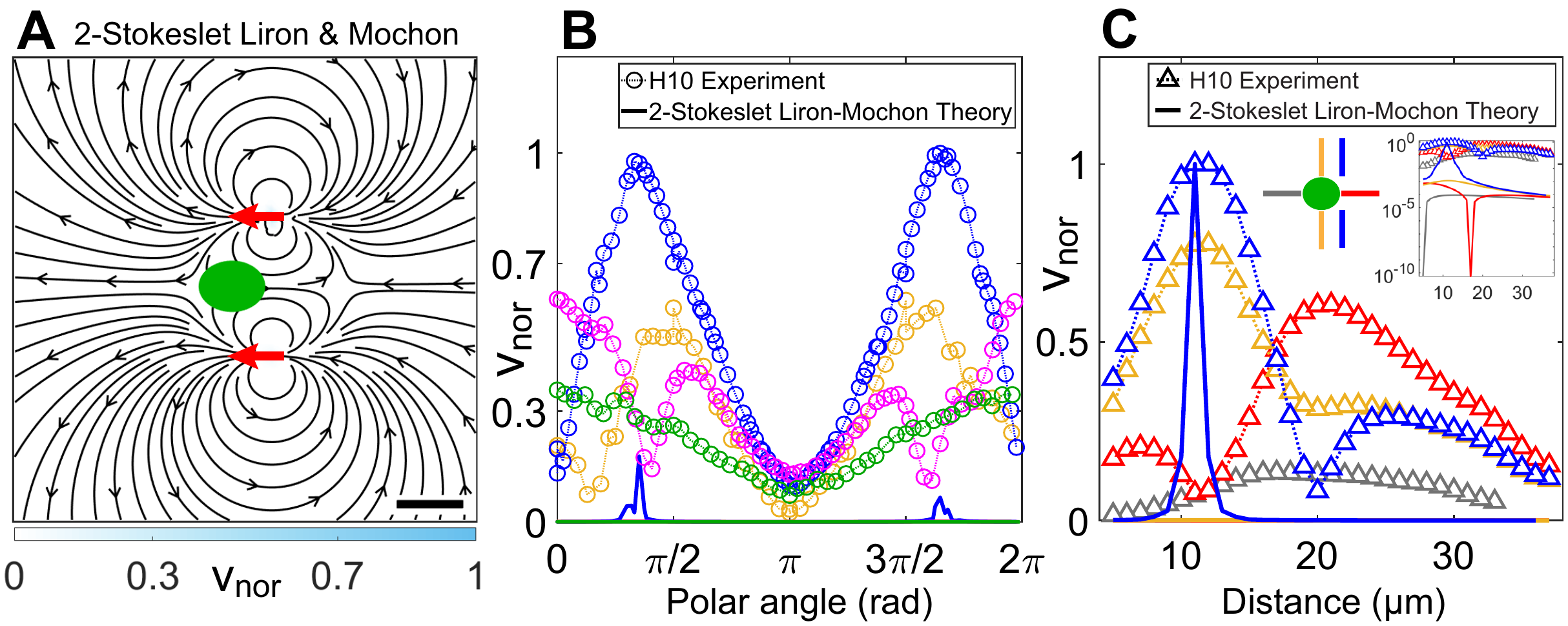}
	\caption{\textbf{Theoretically computed flow field in confinement from Liron \& Mochon's formula.} (\textbf{A}) Theoretically computed flow field using Liron \& Mochon's solution for 2-Stokeslet model in 2D. The red arrows at (6,$ \pm $11)\thinspace\si{\um}  denote the position of the Stokeslets. The colorbar represents flow magnitude normalised by its maximum, $v_{nor} $. Scale bar, \SI{10}{\um}. (\textbf{B}) Comparison between normalised experimental flow of a cell swimming in $ H= \SI{10}{\um}$  (\autoref{fig:Fig3_ExptFlow}C) and Liron \& Mochon's theoretical flow field (A) along representative radial distances, $ r $, from the cell centre as a function of polar angle; $ r~(\si{\um}) = 7$ (yellow), 13 (blue), 20 (magenta), 30 (green). (\textbf{C})~Flow magnitude variation along 4 directions as indicated by separate colors in the \textit{middle} inset [lateral to vortex (blue), lateral to cell centre (yellow), anterior (red), posterior (grey)] for the normalised experimental (symbols) and theoretical (solid lines) velocity fields in \autoref{fig:Fig3_ExptFlow}C and \ref{fig:FigS_A1_LironMochon}A, respectively. Except for the theoretical speed along the vortex direction (blue), others are negligible compared to the experiment as shown in the \textit{rightmost} inset, which is a semilog plot of (C) in the $ y $-axis.}
	\label{fig:FigS_A1_LironMochon} 
\end{figure}

As we have shown that the hydrodynamic cell drag is negligible to the flagellar thrust,  the cell-body drag is insignificant and the observed flow field is mostly due to flagellar thrust.
We, therefore, superpose Liron \& Mochon's solution for two flagellar forces and obtain the flow in \ref{fig:FigS_A1_LironMochon}A. The streamlines of the `\textit{2-Stokeslet Liron \& Mochon flow}' are qualitatively similar to that of the experiment (\autoref{fig:Fig3_ExptFlow}C). However, the 2-Stokeslet theoretical flow of Liron \& Mochon decays much more rapidly than the experimental one and does not capture the experimental flow variation as shown in \ref{fig:FigS_A1_LironMochon}B,C. Notably, there is no signature of vortex position lateral to the forcing point i.e., no minimum in the blue solid curve in \ref{fig:FigS_A1_LironMochon}C because $\boldsymbol{v}^{LM}$ is singular. Therefore, this far-field limit of the theoretical model is insufficient to describe the near-field flow variation, positions of vortices and other flow features of the strongly confined flow accurately. The root mean square deviation (RMSD) in $ v_x $, $ v_y $ \& $ |\boldsymbol{v}| $  between the experimental flow of a H10 cell (\autoref{fig:Fig3_ExptFlow}C) and  2-Stokeslet Liron \& Mochon's flow is 25.9\%, 16.8\% and 30.8\%, respectively (see Materials and Methods for RMSD definition).

\subsection{{\normalsize Inverse Fourier transform of the quasi-2D Brinkman equation in Fourier space}} \label{App_IFT}

The quasi-2D Brinkman equation in Fourier space, \autoref{eq:2DStokesEqnFourier} in the main text, is
\begin{equation} \label{eq:SI2DStokesEqnFourier} 
	\boldsymbol{v}_{\boldsymbol{k}} =\frac{\boldsymbol{O}_{\boldsymbol{k}}\cdot \boldsymbol{F}}{\eta \bigg(k^2 +\dfrac{\pi^2}{H^2} \bigg)}
\end{equation}
Here, the orthogonal projection operator in polar coordinates $ (k,\theta) $ is
\begin{equation}
	\boldsymbol{O}_{\boldsymbol{k}} = 1- \boldsymbol{\widehat{k}}\boldsymbol{\widehat{k}} 
	= \begin{bmatrix}
		1-\widehat{k_x}^2 & -\widehat{k_x}\widehat{k_y}  \\
		-\widehat{k_y}\widehat{k_x} & 1-\widehat{k_y}^2 
	\end{bmatrix}
	=\begin{bmatrix}
		\sin^2\theta & -\sin\theta \cos\theta  \\
		-\sin\theta \cos\theta & \cos^2\theta
	\end{bmatrix}
\end{equation}
where $ \theta $ is the angle between wave vector $ \boldsymbol{k} $ and $ x $-axis. For Stokeslets/Gaussian forces pointing along $ x $- direction only, as in our case, $ \boldsymbol{F} = \begin{bmatrix}
	F   \\
	0
\end{bmatrix} $, therefore $ \boldsymbol{O}(\boldsymbol{k})\cdot \boldsymbol{F} = \begin{bmatrix}
	\sin^2\theta  \\
	-\sin\theta \cos\theta 
\end{bmatrix} F $.

To compute the velocity field in real space, we inverse Fourier transform \autoref{eq:SI2DStokesEqnFourier} in polar coordinates, by replacing the numerator as shown above
\begin{equation}
	\boldsymbol{v}(\boldsymbol{r}) =\frac{1}{(2\pi)^2 \eta} \int e^{i\boldsymbol{k} \cdot \boldsymbol{r}} \begin{bmatrix}
		\sin^2\theta  \\
		-\sin\theta \cos\theta 
	\end{bmatrix}  \thinspace \frac{F ~kdkd\theta}{\bigg(k^2 +\dfrac{\pi^2}{H^2} \bigg)}
\end{equation}
In polar coordinates, the field points in the $ x-y $ plane are given by $ (x,y) = (r\cos \phi, r\sin \phi) $, hence $ \boldsymbol{k} \cdot \boldsymbol{r} = kr\cos(\theta-\phi) $. Thus, the fluid velocity field is
\begin{equation}
	\begin{bmatrix} v_x \\v_y \end{bmatrix} (r,\phi)  =\frac{F}{4\pi^2 \eta} 
	\int_0^{2\pi} d\theta \int_0^\infty dk   \begin{bmatrix} \sin^2\theta \\-\sin\theta \cos\theta \end{bmatrix} \frac{ke^{ikr\cos(\theta-\phi)}}{\bigg(k^2 +\dfrac{\pi^2}{H^2} \bigg)}
\end{equation}
Let us change the $ \theta $ integral from $(0,2\pi) \to (-\pi/2+\phi,\pi/2+\phi) $, where
$ \cos(\theta-\phi)>0 $. For example, the $ \theta $ integral for $ v_x $ changes as follows,
\begin{equation}
	\begin{gathered}
		\int_0^{2\pi} \sin^2\theta e^{ikr\cos(\theta-\phi)} d\theta 
		= \int_{-\frac{\pi}{2}+\phi}^{\frac{\pi}{2}+\phi} \sin^2\theta e^{ikr\cos(\theta-\phi)} d\theta +\int_{\frac{\pi}{2}+\phi}^{\frac{3\pi}{2}+\phi} \sin^2\theta e^{ikr\cos(\theta-\phi)} d\theta
	\end{gathered}
\end{equation} 
Replacing $ \theta \to \theta-\pi $ in the 2nd integral, the limits change as $ (\pi/2+\phi,3\pi/2+\phi) \to (-\pi/2+\phi,\pi/2+\phi)  $, and the integrands $ \sin \theta \to -\sin \theta $, $ \cos \theta \to -\cos\theta $, $ \cos(\theta-\phi) \to -\cos(\theta-\phi) $. Therefore, the 2nd integral in the above equation changes to $ \int_{-\frac{\pi}{2}+\phi}^{\frac{\pi}{2}+\phi} \sin^2\theta e^{-ikr\cos(\theta-\phi)} d\theta $. Hence, $ v_x $'s $ \theta $ integral becomes
\begin{equation}
	\int_0^{2\pi} \sin^2\theta e^{ikr\cos(\theta-\phi)} d\theta = 2 \int_{-\frac{\pi}{2}+\phi}^{\frac{\pi}{2}+\phi} \sin^2\theta \cos[kr\cos(\theta-\phi)] d\theta
\end{equation}
Similarly, $ \int_0^{2\pi} -\sin\theta \cos\theta e^{ikr\cos(\theta-\phi)} d\theta = 2 \int_{-\frac{\pi}{2}+\phi}^{\frac{\pi}{2}+\phi} -\sin\theta \cos\theta \cos[kr\cos(\theta-\phi)] d\theta $. Thus the velocity field in polar coordinates is given by,
\begin{equation} \label{eq:SIFinal2Stokeslet}
	\begin{bmatrix} v_x \\v_y \end{bmatrix} (r,\phi)  =\frac{F}{2\pi^2 \eta} 
	\int_{-\frac{\pi}{2}+\phi}^{\frac{\pi}{2}+\phi} d\theta \int_0^\infty dk   \begin{bmatrix} \sin^2\theta \\-\sin\theta \cos\theta \end{bmatrix} \frac{k\cos[kr\cos(\theta-\phi)]}{\bigg(k^2 +\dfrac{\pi^2}{H^2} \bigg)}
\end{equation}
For Gaussian forces, the numerator just gets multiplied by $ e^{-k^2\sigma^2/2} $. We perform these 2D integrals in MATLAB for a $ 20 \times 23 $ XY grid, with $ k $ integral ranging from 0 to 100 to obtain the theoretical flow fields in this article .

The above integration takes 3 hours of computational time for 2 Stokeslets whereas it takes only 1 minute to compute the flow field for 2 Gaussian forces of $ \sigma =\SI{5}{\um} $ (\textit{Processor:} Intel i7-4770 CPU with clock speed 3.4 GHz). Hence, we try to write a semi-analytical expression for the case of 2 Stokeslets. Let us consider $ kr\cos(\theta-\phi) = p $ and $ \dfrac{\pi r \cos(\theta-\phi) }{H} = q $. Then the $k- $integral changes from $ \int_0^\infty \dfrac{k\cos[kr\cos(\theta-\phi)]}{(k^2 +\pi^2/H^2 )} dk \to \int_0^\infty \dfrac{p \cos p}{p^2+q^2}\thinspace dp$. We rename this integral as $ I(q) $ and calculate it using the Exponential Integral, $ \text{Ei} $ (Eq. 3.723---5 of \cite{Gradshteyn2007}).
\begin{equation}
	I(q) = \int_0^\infty \frac{p \cos p}{p^2+q^2} \thinspace dp = -\frac{1}{2} 
	\big[ e^{-q} \thinspace \overline{\text{Ei}}(q) + e^q \thinspace \text{Ei}(-q) \big]
\end{equation}
where,
\begin{equation} \label{eq:SIEi1}
	\text{Ei}(q) = -\int_{-q}^{\infty} \frac{e^{-m}}{m} \thinspace dm = \int_{-\infty}^{q}  \frac{e^m}{m} \thinspace dm, \quad \text{for}~~q<0
\end{equation}
and to avoid the singularity for $ q>0 $, it is defined by using the principal value of the integral as
\begin{equation} \label{eq:SIEi2}
	\overline{\text{Ei}}(q) = \int_{-\infty}^{-\epsilon}  \frac{e^m}{m} \thinspace dm + \int_{\epsilon}^{q}  \frac{e^m}{m}  \thinspace dm ,~\text{where $ \epsilon>0 $}, \quad \text{for}~~q>0
\end{equation}
In our case $ q>0 $, so we use \autoref{eq:SIEi1} for calculating $ \text{Ei}(-q) $ and \autoref{eq:SIEi2} for calculating $ \overline{\text{Ei}}(q) $, wherein we use $ \epsilon = 10^{-5} $. So, \autoref{eq:SIFinal2Stokeslet} reduces to
\begin{equation}
	\begin{bmatrix} v_x \\v_y \end{bmatrix} (r,\phi)  =\frac{F}{2\pi^2 \eta} 
	\int_{-\frac{\pi}{2}+\phi}^{\frac{\pi}{2}+\phi} d\theta  \begin{bmatrix} \sin^2\theta \\-\sin\theta \cos\theta \end{bmatrix} I(q)  
\end{equation}
This method computes the flow field for 2 Stokeslets in 12 minutes on the same processor.

\subsection{{\normalsize Swimmer based P\'eclet number}} \label{App_Peclet}

Generally, speed and length scales in the definition of P\'eclet number are given by the swimmer speed, $ u $, and radius, $ R $ which we refer to as the swimmer based P\'eclet number, $ Pe_c=uR/D_S $. By this definition,  $ Pe_c^{30} \approx 0.6 $  and $ Pe_c^{10} =0.02 $ for the weakly and strongly confined CR, respectively. However, we note that the flow field closer to the cell surface is dominated by the vortices lateral to the cell body (\autoref{fig:Fig3_ExptFlow}A,C), whose magnitude is significantly higher than the swimmer speed for the strongly confined cell ($ V/u \sim 11 $), in contrast to that of the weakly confined cell ($ V/u \sim 0.3 $). Hence, the flow based P\'eclet number is more appropriate for describing the enhancement of mass transport of solutes due to the vortical flow fields generated by the flagella, particularly for the strongly confined cell ($ H = \SI{10}{\um} $). This is shown below (\ref{tab:TableA1FlowPecletNo}) to be 100 times higher than the swimmer based P\'eclet number, whereas both definitions yield almost similar $ Pe $ for the weakly confined cell ($ H = \SI{30}{\um}$).

\begin{table}[h]
	\begin{center}
	\begin{tabular}{|c|c|c|}
		\hline 
		& $ H = \SI{30}{\um}$ & $ H = \SI{10}{\um}$ \\
		\hline \hline
		Vortical flow speed, $ V $ (\si{\um\per\s}) & 30 & 45 \\
		(\autoref{fig:Fig3_ExptFlow}A,C) & & (also frontal flow) \\
		\hline
		Vortical diameter, $ l_V $ (\si{\um}) & $2 \times 8.5 = 17 $ & $2 \times 20 = 40 $ \\
		2 $ \times $ vortex point distance (\autoref{fig:Fig3_ExptFlow}B,D) & & \\
		\hline
		$ t_{\text{adv}} = l_V/V $ (s)			& 0.57 	& 0.8 \\
		\hline
		$ t_{\text{diff}} = l_V^2/D_S $ (s)		 & 0.3 	& 1.6 \\
		\hline
		$ Pe = t_{\text{diff}}/t_{\text{adv}} = l_V V/D_S $  & 0.5 	& 2 \\
		\hline
	\end{tabular}
	\caption{Flow based P\'eclet number calculation from the flow fields.}
	\label{tab:TableA1FlowPecletNo}
	\end{center}
\end{table}

\subsection{{\normalsize Comparison of our theoretical model of strongly confined flow with that of Jeanneret et al. \cite{Polin2019PRLConfined}}} \label{App_CompareJeanneret}

Jeanneret et al. provides an effective force-free 2D model for explaining the flow field of confined swimmers between 2 boundaries. They consider a force-free combination of 2D Brinkman Stokeslets along with a 2D source dipole to explain their experimental flows \cite{Polin2019PRLConfined}. They use the analytical solution of \cite{PushkinBees2016Brinkman} for their 2D Stokeslets with the permeability length $ \lambda =H/\sqrt{12} $ (for the $ z- $averaged flow in a Hele-Shaw cell of height $ H $). They consider the conventional 3-Stokeslet model of CR where the flagellar thrust, distributed between 2 Stokeslets of strength $ -F_S/2 $ each at  $ (x_1,\pm y_1) $, is balanced by the cell drag of strength $ F_S $ at $ (x_0,0) $, all oriented along the direction of motion. Along with these force-free Stokeslets, they include the 2D source dipole of strength $ I_d $ at $ (x_d,0) $. Finally, they used this model with 6 free parameters $ (F_S,x_0,x_1,y_1,I_d,x_d) $ to fit their experimentally observed flow fields of CR in confinements ranging from 14 to 60\thinspace \si{\um}.

However, our theoretical model consists of a 2D Brinkman Stokeslet because the strongly confined CR exerts a net force on the fluid due to the presence of strong non-hydrodynamic contact friction from the walls, unlike that of \cite{Polin2019PRLConfined}. This force-monopole is spatially distributed equally at the 2 flagellar positions, each with a Gaussian regularization to describe the strongly confined flow due to the H10 cell. The reason our theoretical approach is not the same as \cite{Polin2019PRLConfined} is because there are two major differences in our experimental observations. First, we observe that the strongly confined H10 flow is mostly due to the flagellar motion with a 96\% reduction in the cell's swimming speed, thanks to the static friction from the walls (compared to H30 cells), leading to the hydrodynamic cell-drag  being nearly absent. This coupling between motility and confinement is not observed by \cite{Polin2019PRLConfined}, likely due to the slightly weak confinement ($ D/H \lesssim 0.7 $) produced by their experimental methodology, where the stresses present in the system are mostly hydrodynamic. It is therefore appropriate for them to use the force-free 3-Stokeslet theoretical model for CR (apart from the source dipole contribution) whereas in our case, the nearly absent hydrodynamic drag experienced by the cell body leads to a monopolar flow with only 2 Stokeslets (like-signed) localized with a Gaussian spread around the approximate flagellar positions. Second, the spinning motion of CR cells is restricted in our strongly confined H10 chambers unlike those in \cite{Polin2019PRLConfined}. They added the extra 2D source dipole in their theoretical model to account for both finite-sized effects of the cell body and spinning motion of the cells (explained in Fig. 1c of \cite{Polin2019PRLConfined}).

\subsection{{\normalsize Is the 2-Gaussian Brinkman model applicable to a collection of strongly confined pullers?}} \label{App_2Chlamy}

\renewcommand{\thefigure}{Appendix 1---figure~2}
\begin{figure}[h]
	\centering
	\includegraphics[width=0.75\linewidth]{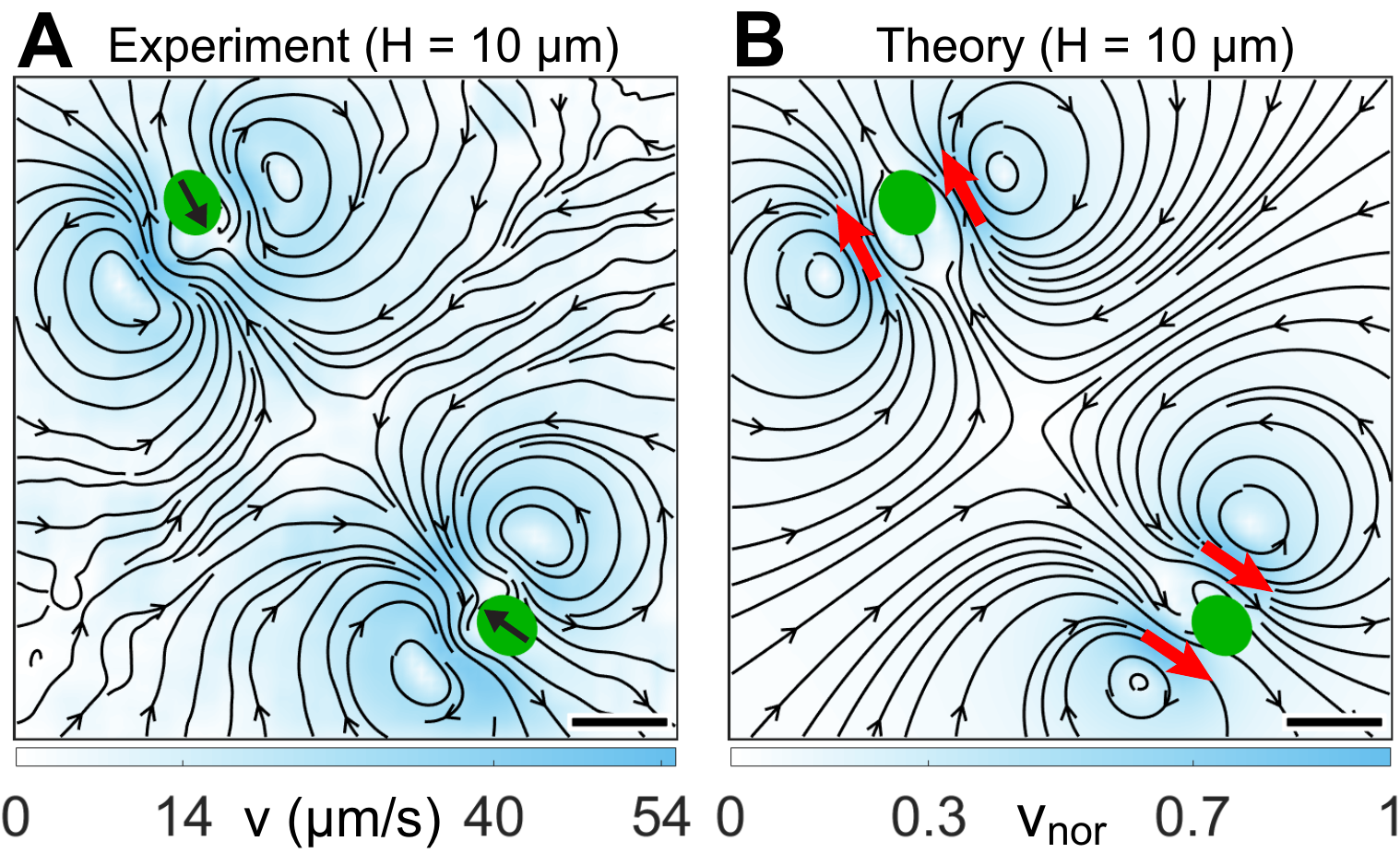}
	\caption{\textbf{Flow fields due to two strongly confined H10 cells.} (\textbf{A}) Experimentally measured flow field for two synchronous cells swimming in $ H= \SI{10}{\um}$. This flow is averaged over $ \sim $ 30 beat cycles for each cell during which the cells move merely $ 0.05  $ times their respective body diameters ($ D \sim \SI{12.42}{\um} $). The centre-to-centre distance between the swimmers is $ 8.75D $. Black arrows on the cell bodies indicate their swimming direction. Solid black lines indicate the streamlines of the flow in lab frame. The colorbar represents flow magnitude, $ v $. (\textbf{B}) Theoretically computed flow field by linearly superposing two 2-Gaussian Brinkman flow, one for each cell. The positions of the pair of 2-Gaussian forces at approximate flagellar positions are denoted by red arrows. The streamlines, vortex flows and stagnation point at the centre of the grid match qualitatively with the experimental one (A). The colorbar represents flow magnitudes normalised by its maximum, $v_{nor} $. Scale bars, \SI{20}{\um}.}
	\label{fig:FigS_A3_2H10Cell} 
\end{figure}

We analyze the fluid flow due to two strongly confined H10 Synchronous cells as a preliminary test for determining the applicability of our theoretical methodology to a collection of microswimmers.  Specifically, we measure the beat averaged flow field of two synchronously beating cells which are separated by $\sim 9$ body diameters and approach each other head-on (\ref{fig:FigS_A3_2H10Cell}A). Therefore, we linearly superpose the solution of the quasi-2D Brinkman equation for a pair of 2-Gaussian forces ($ \sigma = \SI{5}{\um} $) at the approximate flagellar positions of the two cells and obtain the resultant flow field (\ref{fig:FigS_A3_2H10Cell}B). The position and direction of flow vortices along with the stagnation point in between the two cells match well between the experiment and theory. This suggests that  linearly superposing 2-Gaussian Brinkman flows might be an adequate description for the flow field of a dilute collection of CRs.

\end{document}